\documentclass[fleqn,usenatbib]{mnras}

\usepackage{newtxtext,newtxmath}

\usepackage[T1]{fontenc}
\usepackage{ae,aecompl}

\usepackage{graphicx}	
\usepackage{amsmath}	
\usepackage{hyperref}

\def\apj{ApJ}

\def\araa{ARA\&A}
\def\aap{A\&A}
\def\apjl{ApJL}
\def\apjs{ApJS}
\def\mnras{MNRAS}
\def\aj{AJ}
\def\nat{Nature}

\def\icarus{Icarus}

\hypersetup{
  draft=false,
	pdftoolbar=false,
	pdfmenubar=true,
	pdfstartview={FitP},
	breaklinks=false,
	frenchlinks=false,
  colorlinks=true,
  linkcolor=red,
  citecolor=blue,
  filecolor=cyan,
  urlcolor=magenta,
  pdftitle={},
	pdfsubject={},
	pdfauthor={Elspeth K. H. Lee}
  }

\title[3D WD-BD GCMs]{Sunbathing under white light - 3D modelling of brown dwarf - white dwarf atmospheres with strong UV irradiation}

\author[Lee et al.]{
Elspeth K. H. Lee$^{1}$,
Joshua D. Lothringer$^{2}$,
Sarah L. Casewell$^{3}$,
Daniel Kitzmann$^{1}$, \newauthor
Ben W. P. Lew$^{4}$ and
Yifan Zhou$^{5}$
\\
\\
$^{1}$Center for Space and Habitability, University of Bern, Gesellschaftsstrasse 6, CH-3012 Bern, Switzerland \\
$^{2}$Department of Physics, Utah Valley University, Orem, UT 84058, USA \\
$^{3}$School of Physics and Astronomy, University of Leicester, University Road, Leicester LE1 7RH, UK \\
$^{4}$Bay Area Environmental Research Institute and NASA Ames Research Center, Moffett Field, CA 94035, USA \\
$^{5}$Department of Astronomy/McDonald Observatory, The University of Texas at Austin, 2515 Speedway, Austin, TX 78712, USA
}

\date{Accepted XXX; Received YYY; in original form ZZZ}

\pubyear{2022}

\begin{document}
\label{firstpage}
\pagerange{\pageref{firstpage}--\pageref{lastpage}}
\maketitle

\begin{abstract}
The atmospheres of brown dwarfs orbiting in close proximity to their parent white dwarf represent some of the most extreme irradiated environments known.
Understanding their complex dynamical mechanisms pushes the limits of theoretical and modelling efforts, making them valuable objets to study to test contemporary understanding of irradiated atmospheres.
We use the Exo-FMS GCM to simulate the brown dwarfs WD0137-349B, SDSS J141126.20+200911.1B and EPIC212235321B, first coupled to a multi-banded grey radiative-transfer scheme then a spectral correlated-k scheme with high temperature opacity tables.
We then post-process the GCM results using gCMCRT to compare to available observational data.
Our GCM models predict strongly temperature inverted atmospheres, spanning many decades in pressure due to impact of UV band heating.
Post-processing of our models suggest that the day-night contrast is too small in the GCM results.
We therefore suggest that the formation of cloud particles as well as atmospheric drag effects such as magnetic drag are important considerations in setting the day-night temperature contrast for these objects.
\end{abstract}




\section{Introduction}
\label{sec:intro}

To date there are 10 brown dwarfs that are known to be in close (orbital periods $\sim$hrs) orbits around white dwarfs.
These are post-common envelope systems where the brown dwarf survived the formation of the white dwarf, and spiralled inwards within the giant's envelope, resulting in the close, tidally locked orbit we see today.
Despite only 10 of these objects discovered so far, their unique circumstances are highly analogous to hot Jupiter (HJ) systems.
\citet{Steele2011} estimate that only $\approx$ 0.5$\%$ of WDs contain a BD companion, making these systems a rare look into irradiated atmospheric physics at the extremes.
Their short orbital periods also give observational instruments many full orbits of these objects for little time allocation.
Recently, the Jupiter mass planet WD 1856+534 b was discovered orbiting a white dwarf by \citet{Vanderburg2020}, although this object is much cooler than our considered WD-BD pairs, orbiting at $\approx$0.02 AU away from the host star with an equilibirum temperature of $\approx$163 K.
It's orbital period is $\approx$1.41 days, much longer than the orbital periods of the close orbit BD-WD pairs.
In addition, the calculations \citet{Vanderburg2020} suggest that WD 1856+534 b probably didn't evolve to it's current state from post-common envelope interactions unlike the WD-BD systems.

Due to their system properties, close WD-BD binaries occupy the `fourth' temperature corner of sub-stellar objects, namely high irradiation and high internal temperatures \citep{Showman2016}, with typically fast rotation rates of $\approx$ 2 hours, assuming tidal locking.
The WDs in these systems can be extremely hot, with effective temperatures typically T$_{\rm eff}$ $\gtrsim$ 10000-20000 K, with the majority of their emission peaking in the FUV and NUV wavelength ranges.
However, the small radius of the WD means the BDs have typical equilibirum temperatures in the range of the main sequence hot Jupiter population (T$_{\rm eq}$ $\approx$ 1000-2000K).
This makes them useful laboratories for atmospheric theories and modelling in an extreme environment very different but yet familiar regime to HJ atmospheric explorations.

In this paper we examine the WD-BD close orbiting pairs WD0137-349B (WD0137B), SDSS J141126.20+200911.1B (SDSS 1411B) and EPIC212235321B (EPIC2122B).
Table \ref{tab:sys_param} provides the system parameters for all three objects.

\begin{table*}
\centering
\caption{Summary of the WD-BD system parameters used in this study.}
\begin{tabular}{c c c c c c c c c l }  \hline \hline
  Object & T$_{\rm eff, WD}$ & R$_{\rm WD}$ & M$_{\rm BD}$ & R$_{\rm BD}$ & a & P$_{\rm orb}$ & inc. & dist. & refs.  \\
      -   &   [K]             &  [R$_{\odot}$]& [M$_{\rm Jup}$] & [R$_{\rm Jup}$] & [R$_{\odot}$] & [min] & [$^{\circ}$] & [pc] & -  \\ \hline
  WD0137B & 16500 & 0.0186 & 53 & 0.973 & 0.65 & 116 & 35 & 102 &  \citet{Maxted2006} \\
  SDSS 1411B & 13000 & 0.0142 &  52.4 & 0.7  & 0.68 & 121.73 &  90 & 190 & \citet{Littlefair2014} \\
  EPIC2122B & 24900 & 0.017 & 58  & 0.947 & 0.44 & 68.2  &  56 & 386.8 &  \citet{Casewell2018} \\
\hline \hline
\end{tabular}
\label{tab:sys_param}
\end{table*}

WD0137B was discovered by \citet{Maxted2006}, with infrared excess confirming the BD in \citet{Burleigh2006} and in archival 2MASS data \citep{Skrutskie2006}.
\citet{Casewell2015} performed an expansive photometric observation campaign, spanning the V band to the Spitzer 8$\mu$m band, presenting detailed phase curve data of the BD atmosphere.
\citet{Longstaff2017} then detected several atomic species using high-resolution spectroscopy, finding compositional variations between the day and nightside of the BD.

SDSS 1411B was discovered by \citet{Beuermann2013}, finding that the BD eclipses its host WD star.
\citet{Littlefair2014} present photometric observations along with X-Shooter spectroscopic data of the BD.
\citet{Casewell2018b} then present HAWK-I photometric observations of the BD, detecting the nightside of the BD in the J and H bands.

EPIC2122B was discovered by \citet{Casewell2018}, which was the first WD-BD with an orbital period less than 70 minutes.
Archival photometric data from Galex, VST ATLAS, and the VISTA VHS show a clear infrared excess for the WD+BD system.
All three BD have had recent HST WFC3 phase curve measurements published by \citet{Lew2022} (SDSS 1411B) and \citet{Zhou2022} (WD0137B and EPIC2122B), revealing in detail how the BDs near-IR spectra evolve with phase.

In \citet{Lothringer2020}, 1D radiative-convective equilibirum (RCE) modelling was performed using the PHOENIX model for WD0137B and EPIC2122B, showing that the strong UV irradiation was important in setting up strong upper atmospheric temperature inversions.
The also performed retrieval modelling on the objects, finding that non-inverted profiles could fit the data, suggesting that the inclination of these objects plays an important role in what we observe of these atmospheres.
\citet{Tan2020} investigate in-depth the dynamics of WD-BD objects by varying the rotation rate, radiative timescale and drag timescales, using a Newtonian cooling approach.
Their models show how the dynamical regime and number of latitude jet patterns change with rotation and radiative timescales.
\citet{Lee2020} examine the dynamics and radiative properties of the brown dwarf WD0137B using a semi-grey RT approach, comparing the results to the \citet{Casewell2015} data.
They found good agreement with the shape of phase curve data, but over predicted the absolute flux in each band by up to 3x times.
\citet{Sainsbury-Martinez2021} use the DYNAMICO GCM with a Newtonian cooling scheme to investigate the deep atmospheric properties of brown dwarfs, including SDSS 1411B.
Their results suggest that the downward mixing in  SDSS 1411B is weaker compared to main sequence irradiated brown dwarfs.
\citet{Zhou2021} performed GCM models of WD0137B and EPIC 2122B to accompany HST WFC3 data analysis, similar to the \citet{Tan2020} set-up.
This used a simplified semi-grey approach with two UV-Optical shortwaves channels tuned to the 1D models of \citet{Lothringer2020}.
Their GCM models were able to fit the shape of the WFC3 phase curves well.
All of the above studies elucidate the unique temperature and dynamical structures present in these atmospheres, showing that these objects occupy a different dynamical regime to the general hot Jupiter population.

In this study, we expand and improve upon the initial \citet{Lee2020} model by including an ultraviolet (UV) heating component and a deep atmosphere setup.
We then focus on modelling WD0137B, SDSS 1411B and EPIC2122B.
These GCM models are then post-processed in 3D using the gCMCRT code \citep{Lee2021b} to produce observational spectra, and compared to available observational data.

In section \ref{sec:GCM_setup} we detail the GCM set-up for each BD including our RT and opacity schemes.
In section \ref{sec:GCM_results} we show the results of our GCM models.
In section \ref{sec:PP_res} our results from the post-processing of the GCMs is presented.
Section \ref{sec:discussion} presents discussion of our modelling efforts.
Section \ref{sec:conclusions} summarises our project and presents our conclusions.

\section{GCM setup}
\label{sec:GCM_setup}

\begin{table*}
\centering
\caption{Adopted GCM simulation parameters for the WD0137B$^{\bowtie}$, SDSS 1411B$^{\dagger}$ and EPIC2122B$^{\ddagger}$ GCM models.
For T$_{\rm irr}$ we use the \citet{Lothringer2020} values, which are for global heat redistribution efficiencies.}
\begin{tabular}{c c c c c l}  \hline \hline
  Quantity & ${\bowtie}$ & ${\dagger}$ &  ${\ddagger}$ &  Unit & Description \\ \hline
 T$_{\rm irr}$ & 1990 & 1330 & 3435 & K & Stellar irradiation constant \\
 T$_{\rm int}$ & 1400 &  1500 & 1500 & K & Internal temperature \\
 P$_{\rm 0}$ & 1000 &  1000 & 1000 & bar & Reference surface pressure \\
 c$_{\rm P, rad}$ & 14000 & 13600 & 29000 &  J K$^{-1}$ kg$^{-1}$ & Radiative specific heat capacity \\
 c$_{\rm P, ad}$ & 20000 &  20100 & 41700 &  J K$^{-1}$ kg$^{-1}$ & Dynamic specific heat capacity \\
 R$_{\rm d}$ & 3599 & 3600 & 3796 & J K$^{-1}$ kg$^{-1}$  & Specific gas constant \\
 $\kappa_{ad}$ & 0.18 & 0.18 & 0.09 &  J K$^{-1}$ kg$^{-1}$  & Adiabatic coefficient \\
 g$_{\rm BD}$ & 1259 & 2630 & 1585  &  m s$^{-2}$ & Acceleration from gravity \\
 R$_{\rm BD}$ & 0.973 & 0.87 & 0.947 &  R$_{\rm Jup}$ & Radius of brown dwarf \\
 $\Omega_{\rm BD}$ & 9.06 $\times$ 10$^{-4}$ & 8.60 $\times$ 10$^{-4}$ & 1.5 $\times$ 10$^{-3}$ & rad s$^{-1}$ & Rotation rate of brown dwarf \\
 $\Delta$t$_{\rm dyn}$ & 10/20 & 10/20 & 10/20 & s  &  Corr-k/Grey Dynamical time-step \\
 $\Delta$t$_{\rm rad}$ & 40 & 40 & 40 & s  &  Corr-k radiative time-step \\
 N$_{\rm v}$ & 54 & 54  & 54 &  - & No. vertical layers \\
 d$_{\rm 2}$ & 0.02 & 0.02  & 0.02  &  - & Div. dampening coefficient \\
\hline \hline
\end{tabular}
\label{tab:GCM_param}
\end{table*}

We use the Exo-FMS GCM in a gas giant configuration \citep[e.g.][]{Lee2021} using a multi-band grey opacity set-up and a correlated-k scheme.
We primarily use the multi-band scheme for the GCM spin-up up to 1000 days, we then run for another 100 days, taking the average of this period as the final `result' of the multi-band GCM models.
We then restart the model at 1000 days but using the correlated-k scheme.
This is run for 200 days, then another 50 days, the average of which is given as the final answer for the correlated-k cases.
Table \ref{tab:GCM_param} gives the GCM parameters for the WD0137B, SDSS 1411B and EPIC2122B simulations respectively.
Our simulations were performed with a cubed-sphere C48 ($\approx$ 192 x 96 in longitude and latitude elements) resolution grid with 54 vertical layers between 10$^{-6}$-1000 bar.

\subsection{Two-stream RT}

We use the same two-stream set-up as the ultra hot Jupiter (UHJ) models in Lee et al. (in prep.) for the columnwise RT in the GCM.
For modelling the irradiation from the white dwarf we use the adding method of \citet{Mendonca2015}, which is an efficient shortwave scheme that includes the effects of scattering.
For the longwave RT inside the GCM we use the short characteristics method of \citet{Olson1987} with linear interpolants, which is a quick, accurate and stable two-stream method for non-scattering atmospheres.
We improve the stability of the radiative-transfer by interpolating the temperature layers to the temperature levels using piecewise quadratic Bezier interpolation following \citet{Hennicker2020}, instead of linear interpolation as in the previous versions.
This interpolation is used in the multi-band and correlated-k opacity modes.
Our two-stream approaches are available on GitHub \footnote{\url{https://github.com/ELeeAstro}}.

\subsection{Multi-band grey scheme}

\begin{table*}
\centering
\caption{Multi-banded grey simulation parameters for the WD0137B, SDSS 1411B and EPIC2122B GCM models.
These values were chosen to best fit the T-p profiles from the \citet{Lothringer2020} 1D RCE models.}
\begin{tabular}{c c c c c}  \hline \hline
 WD0137B & a & b & c & $\beta$ \\ \hline
 UV$_{1}$ & -0.01886987 & 0.1461967 & 2.25388489 &  0.001 \\
 UV$_{2}$ & -0.01536777 & 0.29015277 & 0.1927073 & 0.02 \\
 UV$_{3}$ & 0.01349805 &  0.32146754 &  -1.53954299 & 0.075 \\
 V({$\tau$ = 1 at 1 bar}) &  0.0 & 0.0 & $\log_{10}$(0.01259) & 0.90 \\
 IR & 0.05358019 & -0.1411748 &  -3.23944985 & 1.0 \\ \hline \hline
 SDSS 1411B & a & b & c & $\beta$ \\ \hline
 UV$_{1}$ & 0.01061987 & 0.01009008 & 1.05605325 &  0.1 \\
 UV$_{2}$ & 0.03041112 & 0.19607576 & -1.68413654 & 0.05 \\
 UV$_{3}$({$\tau$ = 1 at 0.01 bar}) & 0.0 & 0.0 & $\log_{10}$(2.63) & 0.05 \\
 V({$\tau$ = 1 at 1 bar}) &  0.0 & 0.0 & $\log_{10}$(0.0263) & 0.05 \\
 IR &  0.06014871 & -0.28306213 & -2.98933553 & 1.0 \\ \hline \hline
 EPIC2122B & a & b & c & $\beta$ \\ \hline
 UV$_{1}$ & -0.01516223 & 0.39847058 & -0.10444337 &  0.05 \\
 UV$_{2}$ & 0.02793481 & 0.21704063 & -1.07090462 & 0.2 \\
 UV$_{3}$ & 0.01539364 & -0.08168898 & -1.61413292 & 0.5 \\
 V$_{2}$({$\tau$ = 1 at 0.1 bar}) &  0.0 & 0.0 & $\log_{10}$(0.1585) & 0.5 \\
 IR & 0.02918899 & -0.11979848 & -1.99231188 & 1.0 \\ \hline \hline
\end{tabular}
\label{tab:UV_param}
\end{table*}

We develop a simplified multi-band scheme, that attempts to capture the UV wavelength absorption in the BD atmosphere in a more realistic way.
For the shortwave UV-Optical (UV-Opt) bands we assume three pressure dependent opacity functions, which are fit to the PHOENIX forward models of \citet{Lothringer2020}.
We produce UV-Opt band opacities by taking the cross-section output from the \citet{Lothringer2020} models to create pressure dependent Planck mean fitting functions.
The 3 bands we chose are generally for the wavelength ranges 0.15-0.3, 0.3-0.4 and 0.4-1.0 $\mu$m, although considerations were made for where the opacity structure changes with wavelength for each of the 1D atmospheres.
The UV-Opt band opacities, $\kappa$ [m$^{2}$ kg$^{-1}$], take the same form as \citet{Tan2019}, given by
\begin{equation}
\label{eq:opac_P}
\log_{10}\kappa = a\log_{10}p^{2} + b\log_{10}p + c,
\end{equation}
where $p$ is the pressure in Pa and a, b, c are fitting coefficients.
We then take a 4th band as a flexible `visual' band, using a constant grey opacity.
The exact wavelength of this band is arbitrary.
This is then fit to the deep inversion region found in the \citet{Lothringer2020} models.
We note that this opacity set-up is not as comprehensive as using temperature dependent opacities, but our scheme captures the gross radiative effect from the UV heating.
A zero Bond albedo is assumed for all UV-Opt bands, instead the fraction of the total incident flux carried by each band is given by the $\beta$ parameter.
We are flexible with fitting the $\beta$ parameters, with the total not necessarily adding up to unity.
Table \ref{tab:UV_param} presents the multi-band parameters for each brown dwarf.

To fit the large upper atmosphere UV heating profiles in the \citet{Lothringer2020} we experiment and adjust the fraction of the irradiated flux in each UV-Opt band until a decent fit to the 1D RCE profile is found.
Each individual WD-BD system requires it's own set of opacity coefficients and band fractions.
As with the UV-Opt bands, we produce a pressure dependent infrared opacity function (Eq. \ref{eq:opac_P}) from the opacity data of \citet{Lothringer2020}.
However, we use only one IR band and use the Rosseland mean instead of the Planck mean \citep[e.g.][]{Heng2017b}.
The Rosseland mean is more appropriate for use in the IR band as it better captures the continuum opacities which are important for setting the level of the IR photospheric regions.

\subsection{Correlated-k scheme}

\begin{table}
\centering
\caption{Opacity sources and references used in the GCM correlated-k scheme and gCMCRT post-processing.
bf = bound-free, ff = free-free.}
\begin{tabular}{c l}  \hline \hline
Opacity Source &   \\ \hline
Atomic &  Reference \\ \hline
Na  &  \citet{Kurucz1995}   \\
K   &  \citet{Kurucz1995}  \\
Fe   &  \citet{Kurucz1995}  \\
Fe$^{+}$   &  \citet{Kurucz1995}  \\ \hline
Molecular &  \\ \hline
H$_{2}$O   &  \citet{Polyansky2018}   \\
OH   &  \citet{Hargreaves2019}   \\
CH$_{4}$   &  \citet{Hargreaves2020}  \\
C$_{2}$H$_{2}$   &  \citet{Chubb2020}   \\
NH$_{3}$   &  \citet{Coles2019}   \\
CO   &  \citet{Li2015}   \\
CO$_{2}$    &  \citet{Yurchenko2020}   \\
PH$_{3}$ & \citet{Sousa_Silva2015} \\
H$_{2}$S   & \citet{Azzam2016} \\
HCl   & \citet{Gordon2017} \\
HCN   & \citet{Harris2006, Barber2014} \\
SH & \citet{Gorman2019} \\
HF & \citet{Li2013,Coxon2015} \\
N$_{2}$ & \citet{Western2018} \\
H$_{2}$ & \citet{Roueff2019} \\
SiO & \citet{Yurchenko2021} \\
TiO & \citet{McKemmish2019} \\
VO & \citet{McKemmish2016} \\
MgH & \citet{GharibNezhad2013} \\
CaH & \citet{Bernath2020} \\
TiH & \citet{Bernath2020} \\
CrH & \citet{Bernath2020} \\
FeH & \citet{Bernath2020} \\ \hline
Continuum  & \\ \hline
H$_{2}$-H$_{2}$  &  \citet{Karman2019} \\
H$_{2}$-He   & \citet{Karman2019} \\
H$_{2}$-H   & \citet{Karman2019} \\
H-He   & \citet{Karman2019} \\
H$^{-}$ (bf/ff) &  \citet{John1988} \\
H$_{2}^{-}$ (ff)  & \citet{Bell1980} \\
He$^{-}$ (ff)  & \citet{Bell1982} \\ \hline
Rayleigh scattering & \\ \hline
H$_{2}$ & \citet{Dalgarno1962} \\
He  & \citet{Thalman2014} \\
H  & \citet{Kurucz1970} \\
e$^{-}$ & Thomson scattering \\
\hline \hline
\end{tabular}
\label{tab:line-lists}
\end{table}

In addition to developing the multi-band grey scheme, we attempt to directly use a correlated-k (corr-k) scheme within the GCMs.
Our motivation to use the corr-k scheme is to see if our current opacity tables (used for UHJs) and species are complete enough for WD-BD modelling and to identify important species missing in our opacity tables.
Furthermore, we are interested in attempting to reproduce the 1D RCE T-p structures using the corr-k scheme, without tuning of opacities as done in the band grey method.

For the spectral bands we use the 32 bands of \citet{Amundsen2016}, which captures important NUV absorption of atoms and molecules out to 0.2 $\mu$m.
We create a pre-mixed k-table with 4+4 g-ordinances as in \citet{Showman2009, Kataria2014}, the same as the UHJ models performed in Lee et al. (submitted).
Table \ref{tab:line-lists} shows the opacities included in our scheme.
For the pre-mixed tables, we assume solar metallicity \citep{Asplund2009} and chemical equilibirum (CE) using ggCHEM \citep{Woitke2018} to weight each species by their volume mixing ratio \citep[e.g.][]{Amundsen2017}.
We include the effects of condensation and thermal decomposition/ionisation of species in making the CE tables.
Our pre-mixed k-table is valid between a temperature of 200-6100 K and pressures between 10$^{-8}$-1000 bar, suitable for the range of temperatures and pressures for WD-BD atmospheric modelling.

In testing with the GCM, we found a dynamical timestep of 10s and a radiative timestep of 40s was required to stabilise the corr-k models effectively.
This leads to a prohibitive computational time to start the simulations from scratch, so we chose to restart the simulations at 1000 days from the results of the multi-band scheme, and run for an additional 250 days, averaging the last 50 days as the final result.
Due to the higher gravity and upper atmosphere temperatures of the irradiated BDs, their radiative-timescales are expected to be several magnitudes shorter compared to HJ population \citep{Lee2020}.
This suggests that radiative-equilibirum may be achieved much quicker when simulating these atmospheres, and to a deeper pressure compared to regular HJ atmospheres.
Additionally, analysis of the total kinetic energy of the GCM with time in \citet{Lee2020} suggest that these types of simulation quickly settle into statistical dynamical equilibirum as well.

The results of tests performed in \citet{Lothringer2020} and \citet{Lothringer2020b} suggest that the main opacity sources in the UV-Opt responsible for temperature inversions in WD-BD systems are TiO, VO, SiO, Fe and Fe$^{+}$.
We therefore include these species in our pre-mixed k-tables.
Importantly, we include pressure broadening of Fe and Fe$^{+}$ lines when calculating their opacity tables due to the increased gravity of brown dwarfs which results in the incident energy deposition occurring at higher pressures compared to HJ systems.

\subsection{Other GCM improvements}

\citet{Carone2020} suggest that the pressure boundaries in GCMs should be increased for fast rotation rates; accordingly we choose a lower boundary of 1000 bar for all simulations.
The visual optical depth can also remain small to the lower boundary due to the high surface gravity of the brown dwarfs \citep{Lee2020}, so increasing the maximum pressure ensures that the absorbed irradiation from the white dwarf is adequately captured.
To help stabilise the simulation, we include a linear Rayleigh `basal' drag \citep{Liu2013, Komacek2016, Tan2019, Carone2020}, which applies between 1000 to 100 bar.
We chose a baseline drag timescale of 2 Earth days.
This drag is physically justified as a deep magnetic drag effect, however, it is unclear how much magnetic drag is appropriate for WD-BD objects.
Recently \citet{Beltz2022} has shown that considering magnetic drag on the atmospheric flow in hot Jupiters can significantly change the dynamical properties of the atmosphere.
For the T-p profile initial conditions, we use the \citet{Parmentier2015} picket fence scheme as discussed in \citet{Lee2021}.

For the GCM heat capacity and specific gas constant we use values that correspond to the deep adiabatic region from the \citet{Lothringer2020}, to attempt to reproduce the 1D models adiabatic gradient in the deep atmosphere inside the GCM.
However, we chose a different heat capacity for the heating from the RT scheme, preferring to take the heat capacity values at $\sim$10 bar to better capture the heating/cooling rates at the IR photospheric regions of the BD.

\section{GCM results}
\label{sec:GCM_results}

In this section, we present the GCM results for each of the three BD models.
We show both the multi-band grey and corr-k results.

\subsection{WD0137B}

\begin{figure*} 
   \centering
   \includegraphics[width=0.49\textwidth]{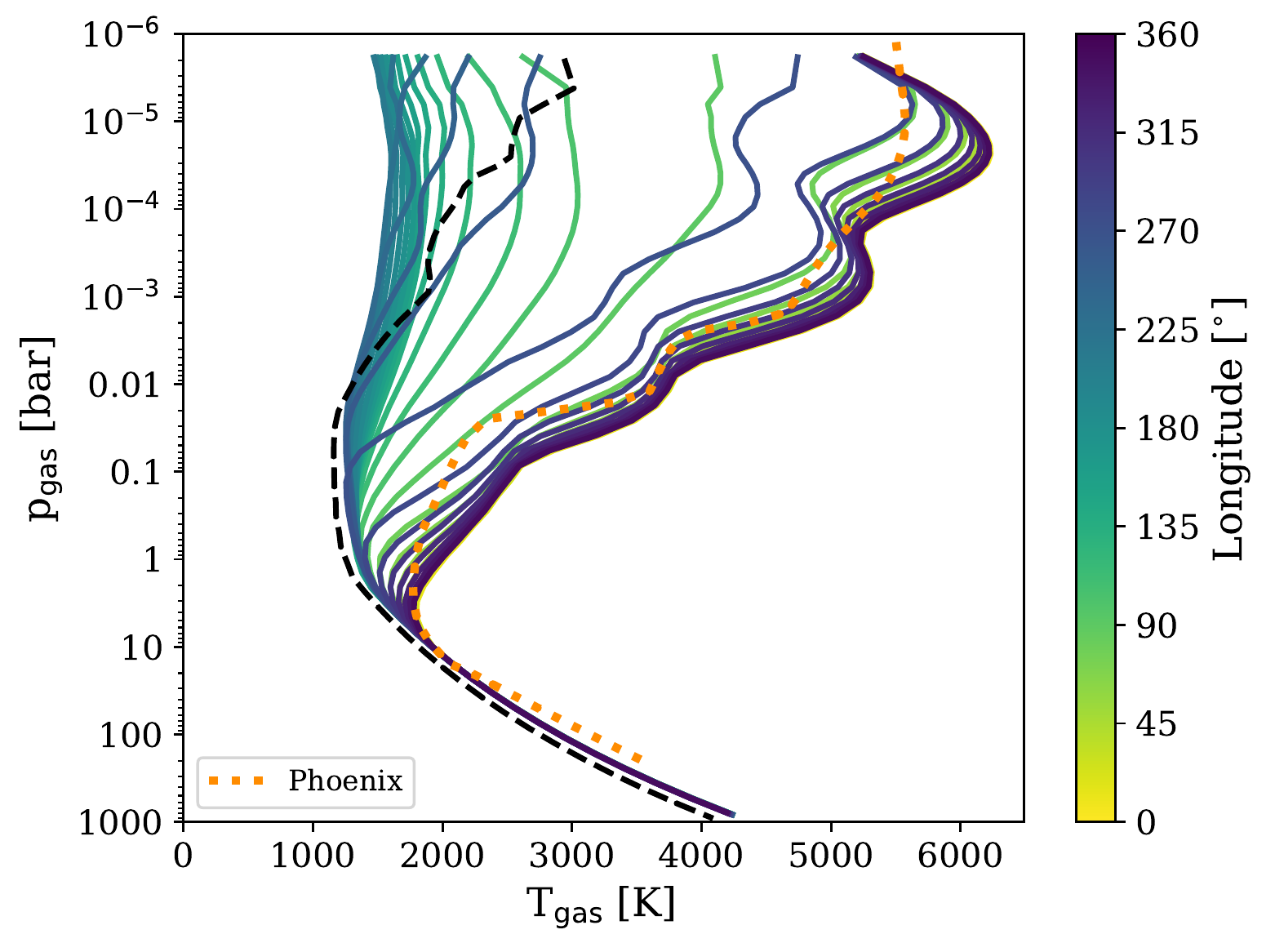}
   \includegraphics[width=0.49\textwidth]{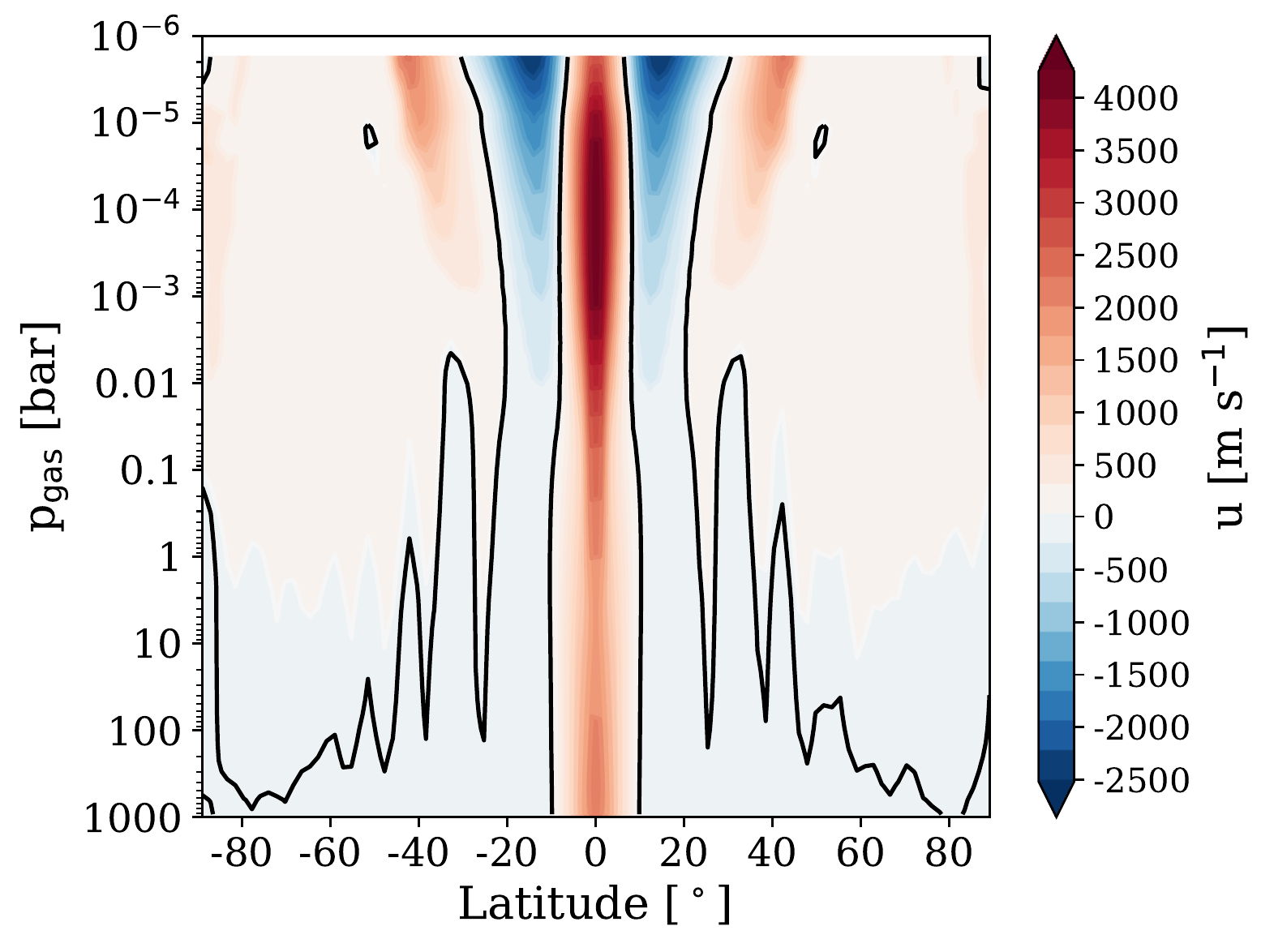}
   \includegraphics[width=0.49\textwidth]{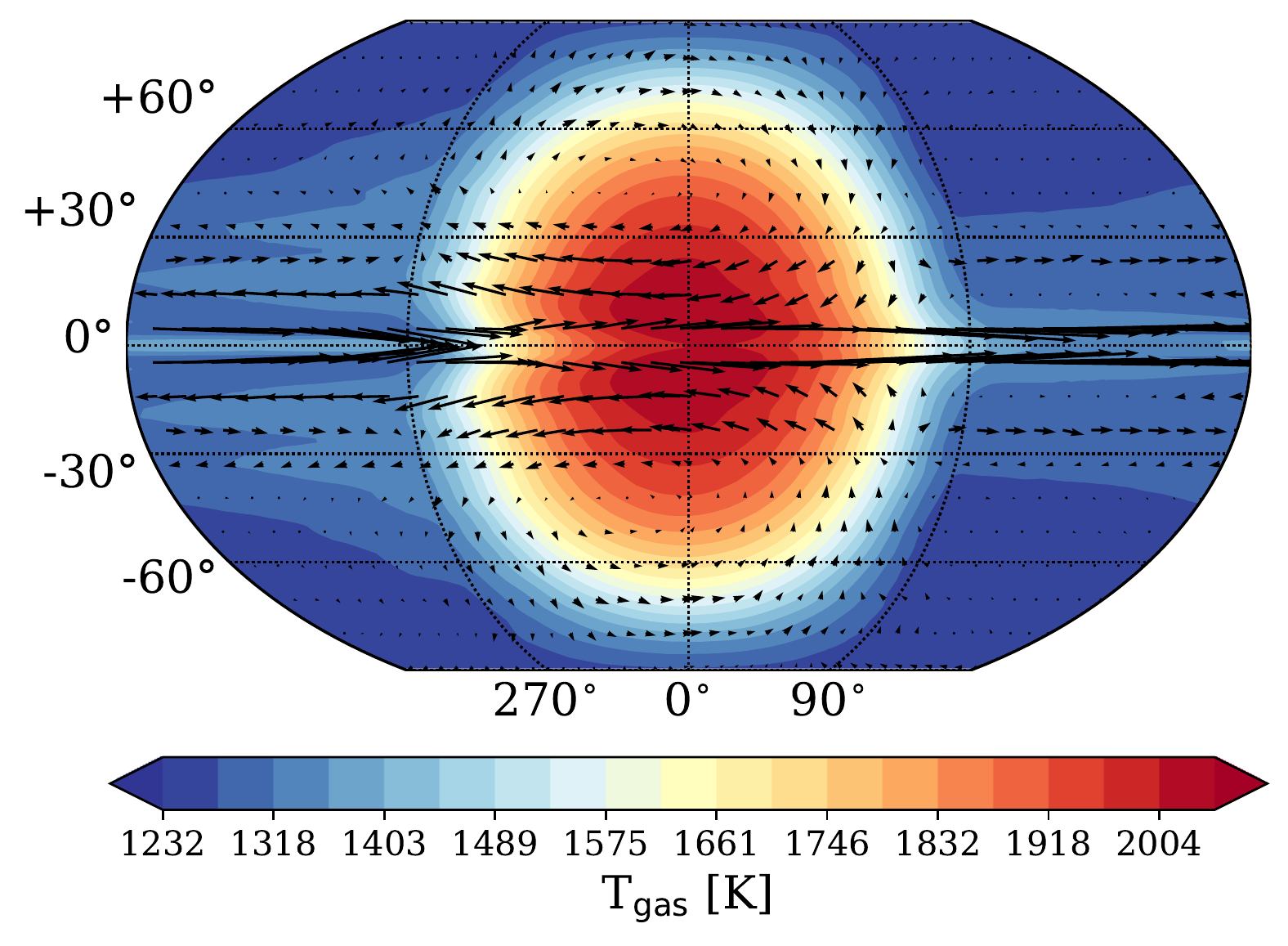}
   \includegraphics[width=0.49\textwidth]{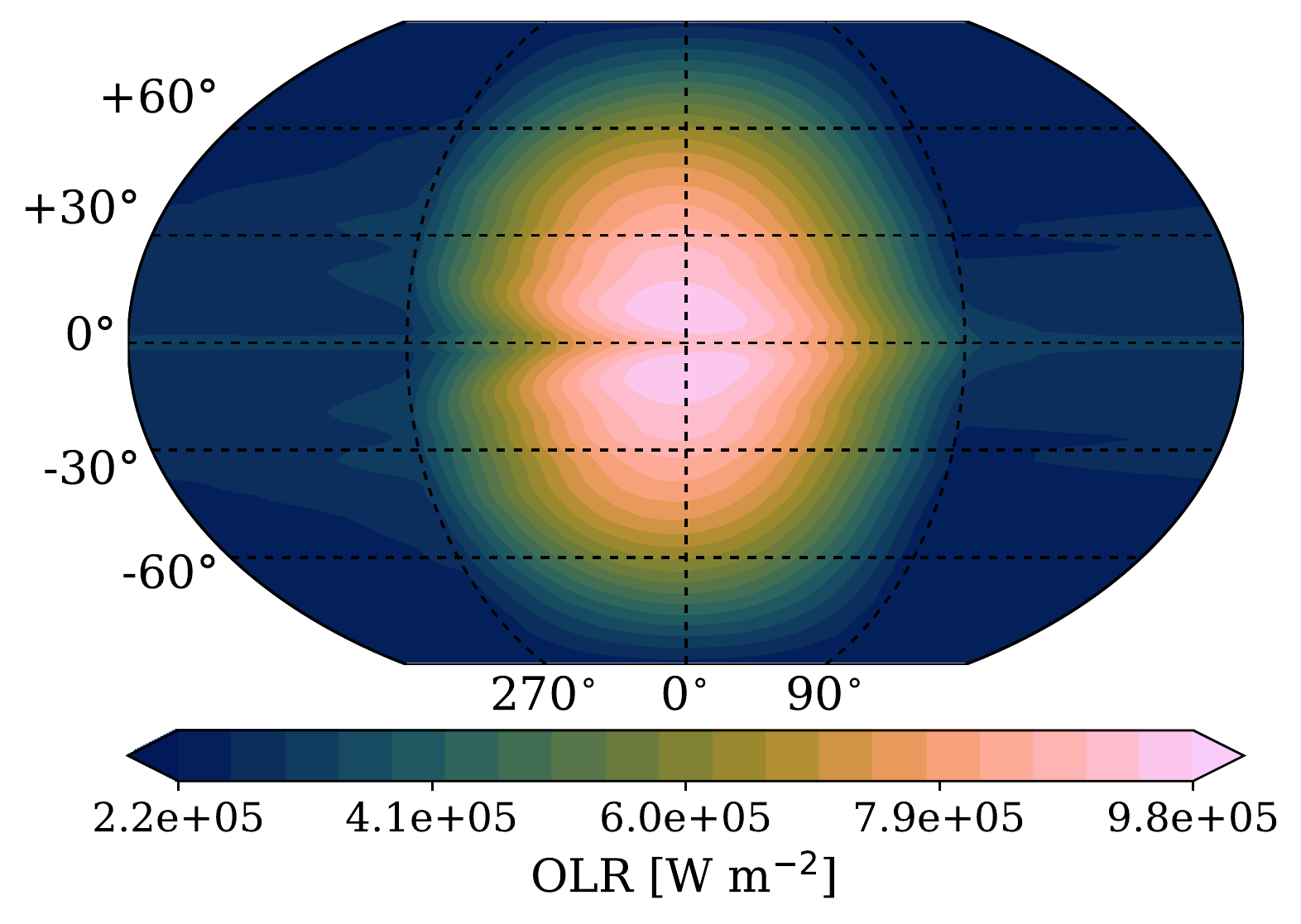}
   \caption{WD0137B GCM results from Exo-FMS using the banded grey opacity scheme.
   Top left: Vertical T-p profiles at the equatorial regions (solid coloured lines) and polar column (dashed).
   Top right: Zonal mean zonal velocity.
   Bottom left: Temperature map at 1 mbar.
   Bottom right: Outgoing longwave radiation (OLR) map.}
   \label{fig:WD0137B_GCM_grey}
\end{figure*}

In figure \ref{fig:WD0137B_GCM_grey} we show the results of the WD0137B GCM using the multi-band grey scheme.
From the T-p profiles it is clear that the multi-band scheme represents well the 1D RCE modelling from \citet{Lothringer2020}, representing both the deep and upper atmosphere structures well.
The zonal mean zonal velocity plot shows a typical BD-WD dynamical regime, with a thin equatorial jet and multiple jets at latitude.
Our temperature and outgoing longwave radiation (OLR) maps again show typical WD-BD behaviour, with hot Rossby lobes on the dayside near the equator \citep{Lee2020, Tan2020}.

\begin{figure*} 
   \centering
   \includegraphics[width=0.49\textwidth]{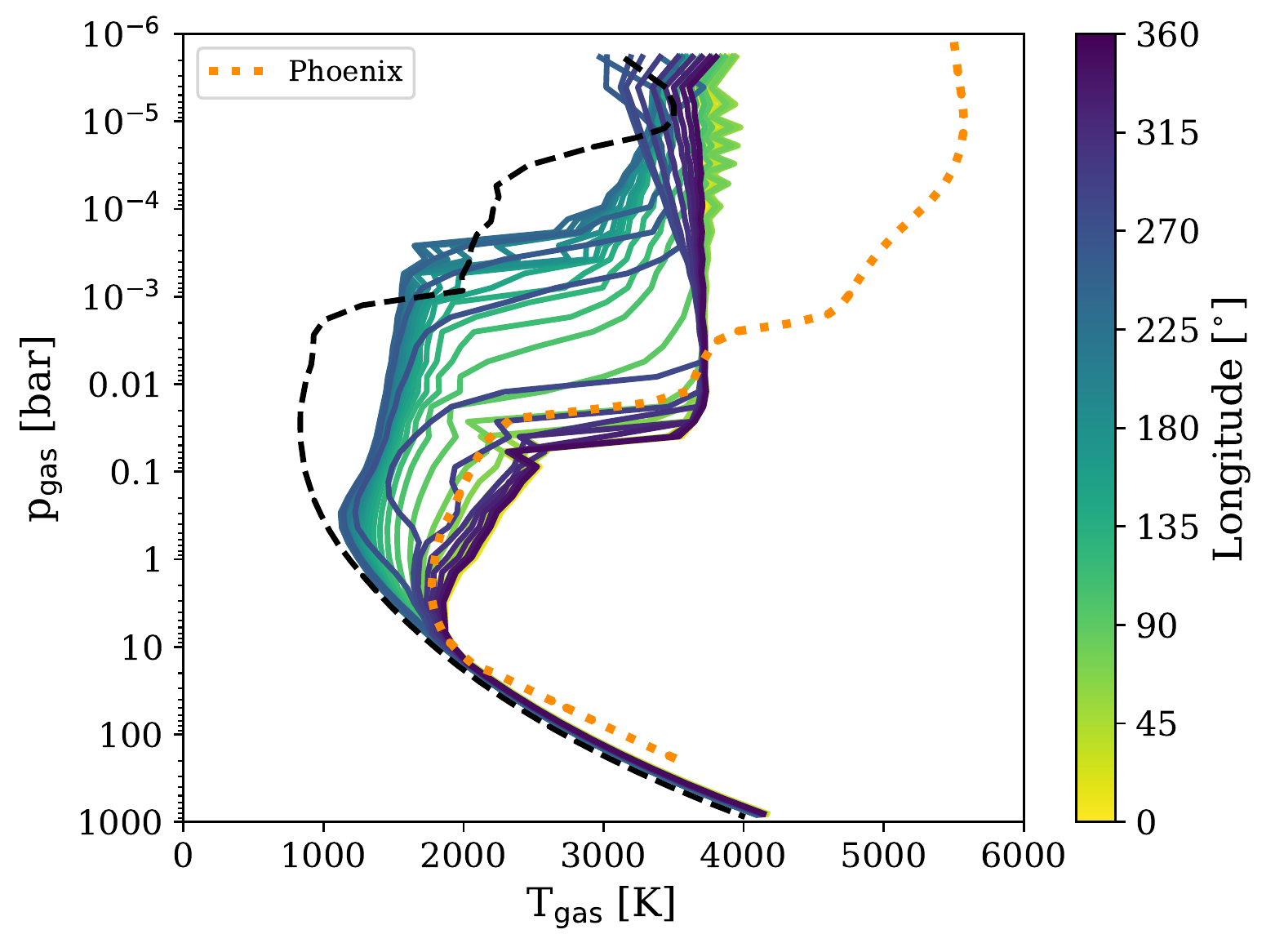}
   \includegraphics[width=0.49\textwidth]{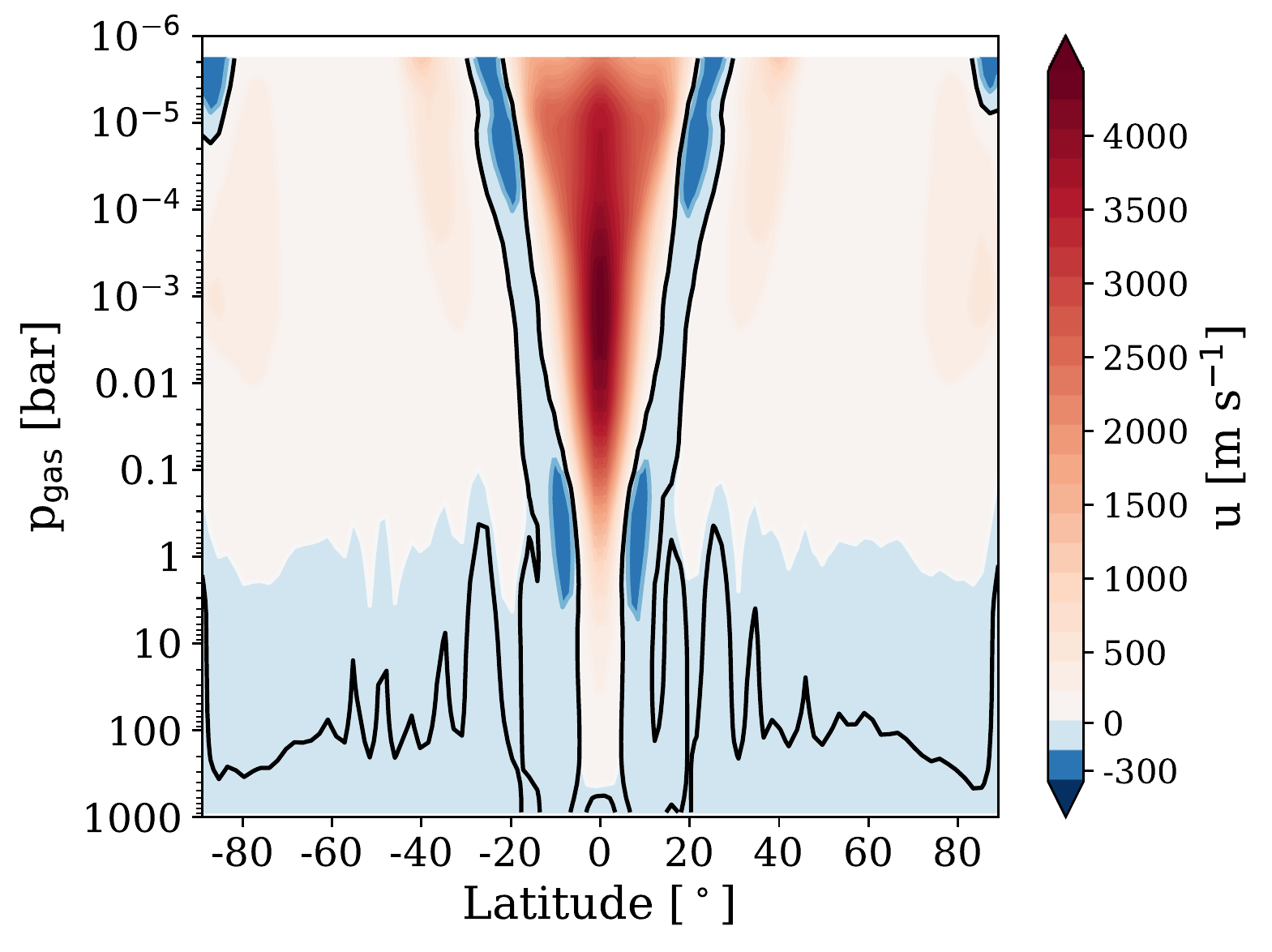}
   \includegraphics[width=0.49\textwidth]{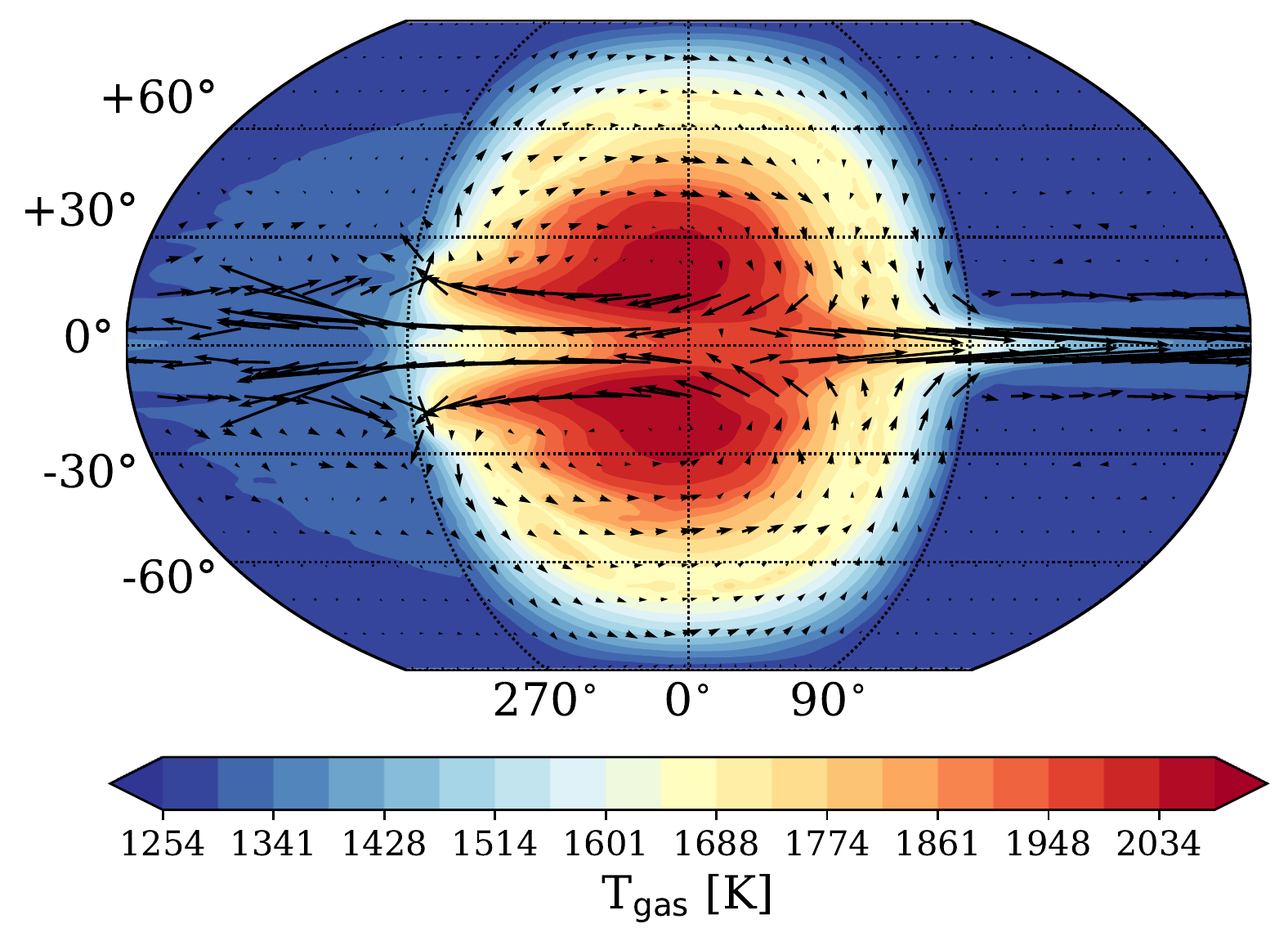}
   \includegraphics[width=0.49\textwidth]{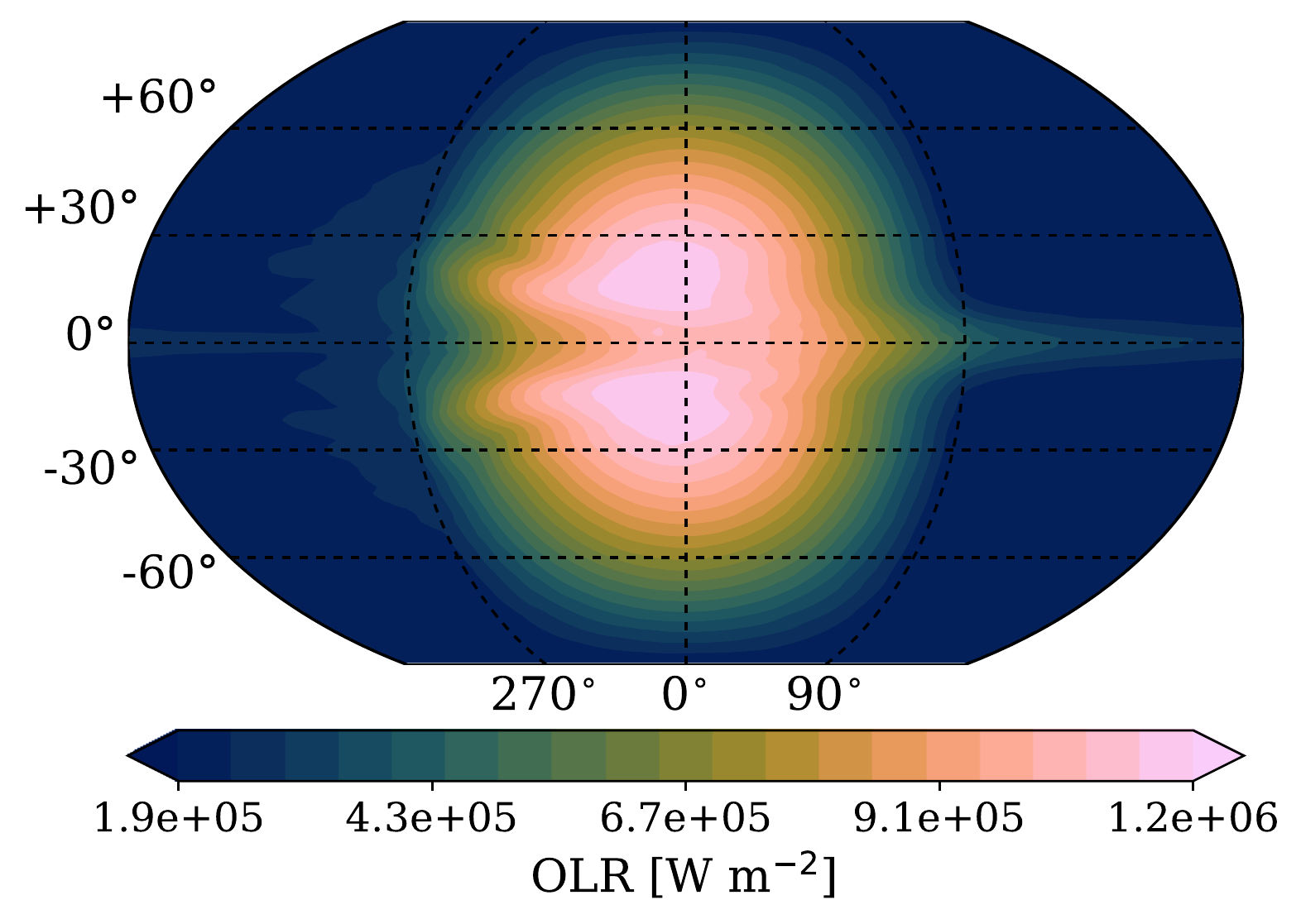}
   \caption{WD0137B GCM results from Exo-FMS using the correlated-k scheme.
   Top left: Vertical T-p profiles at the equatorial regions (solid coloured lines) and polar column (dashed).
   Top right: Zonal mean zonal velocity.
   Bottom left: Temperature map at 1 mbar.
   Bottom right: Outgoing longwave radiation (OLR) map.}
   \label{fig:WD0137B_GCM_ck}
\end{figure*}

The corr-k model results (Fig. \ref{fig:WD0137B_GCM_ck}) show a sharper T-p structure more in line with the 1D model results.
However, the corr-k fails to represent at all the absorption at very low pressures ($<$ 0.01 bar), showing an isothermal profile below this pressure.
This suggests that the corr-k model is not accurately capturing the low pressure UV absorption unlike in the 1D RCE model.
The overall dynamical structures remain similar to the multi-band grey model in this case however.

\subsection{SDSS 1411B}

\begin{figure*} 
   \centering
   \includegraphics[width=0.49\textwidth]{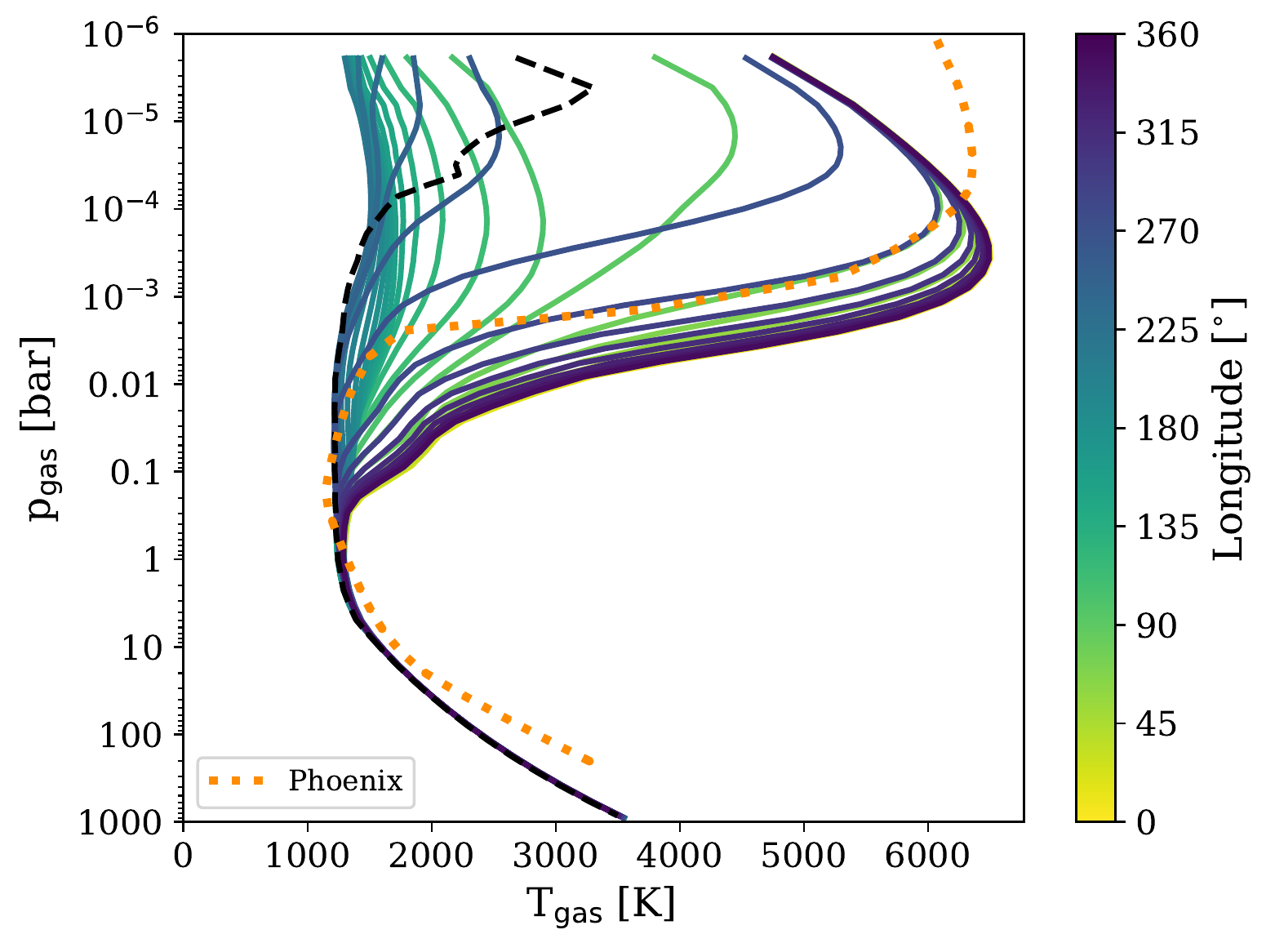}
   \includegraphics[width=0.49\textwidth]{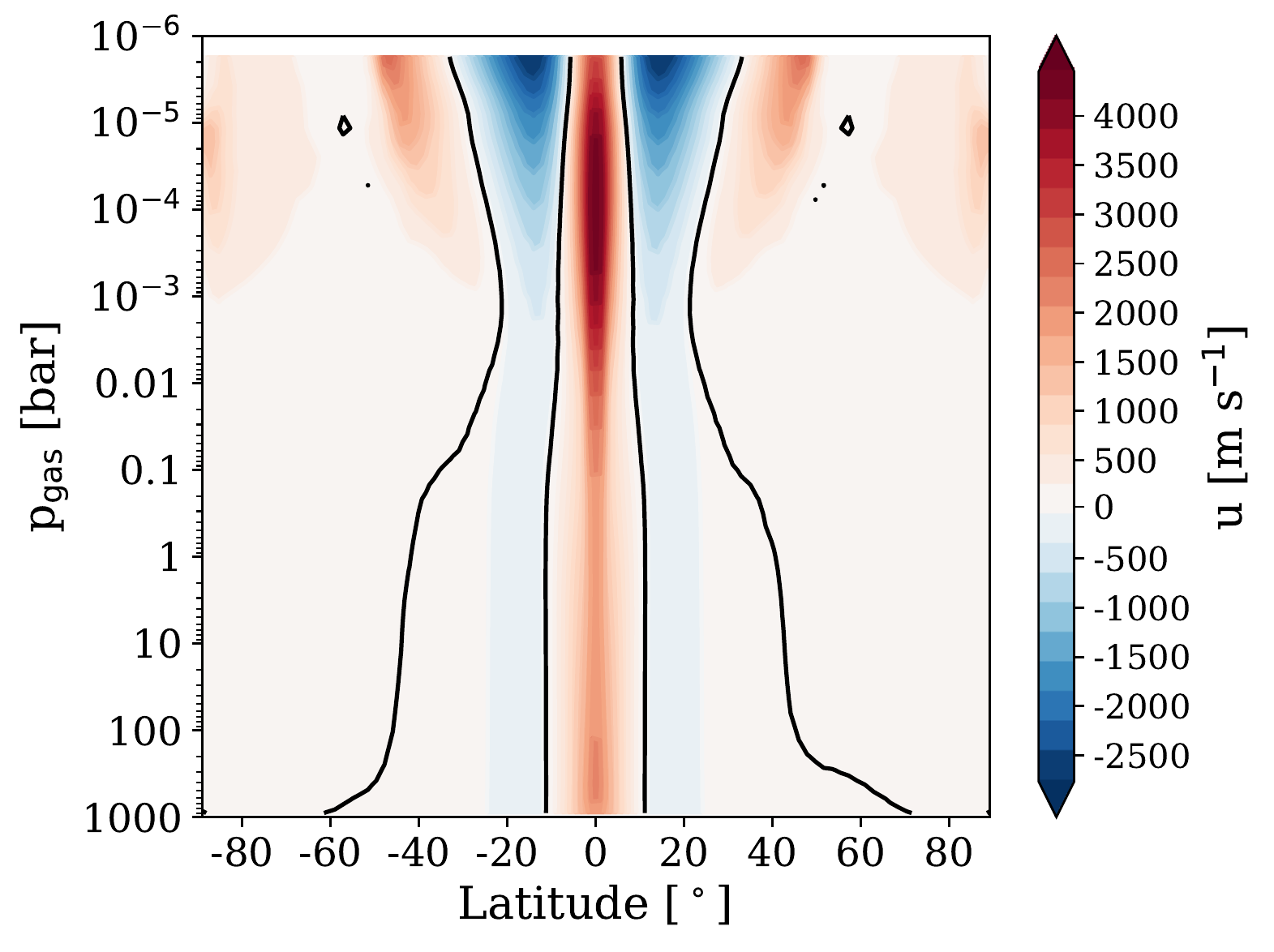}
   \includegraphics[width=0.49\textwidth]{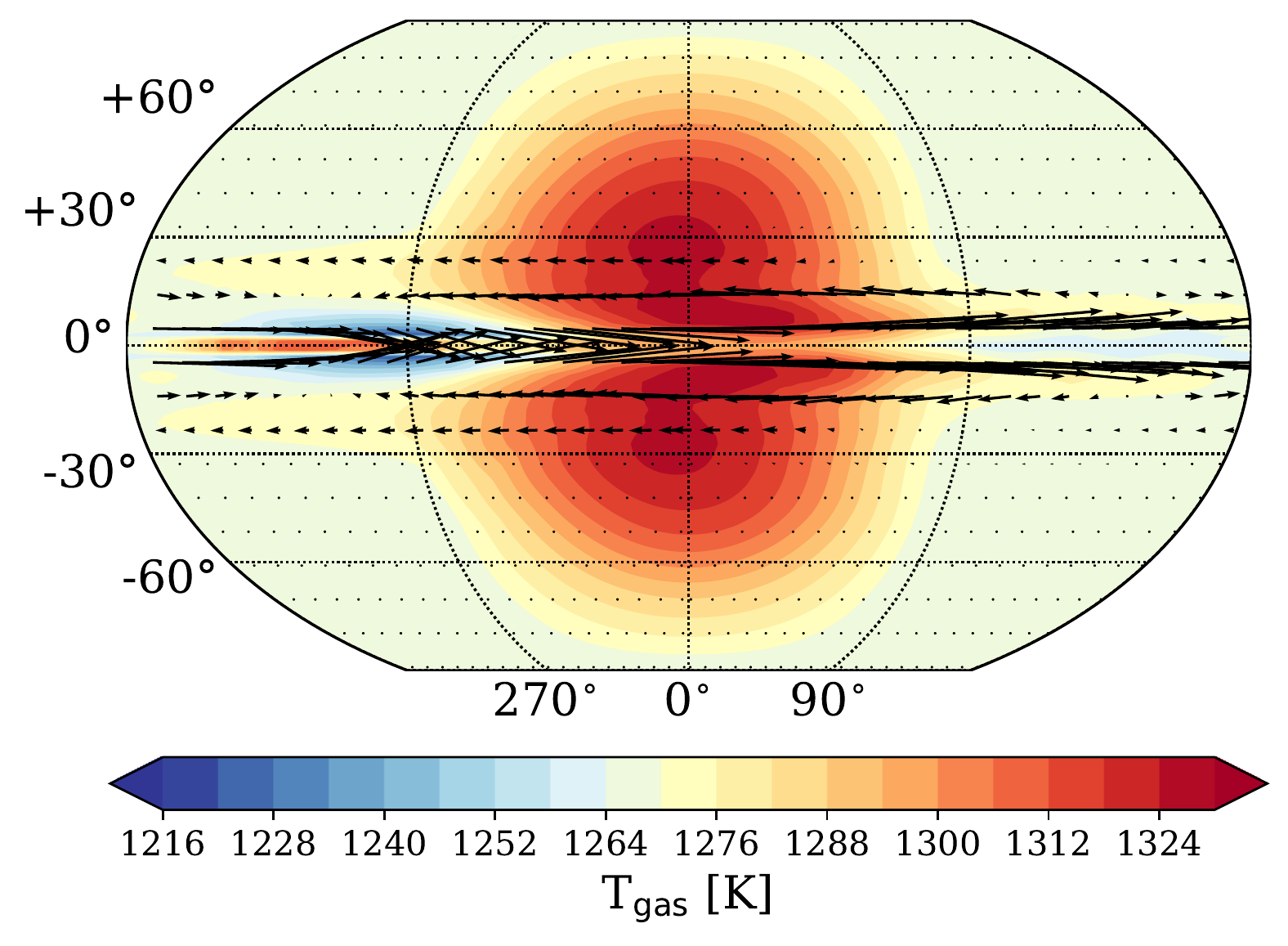}
   \includegraphics[width=0.49\textwidth]{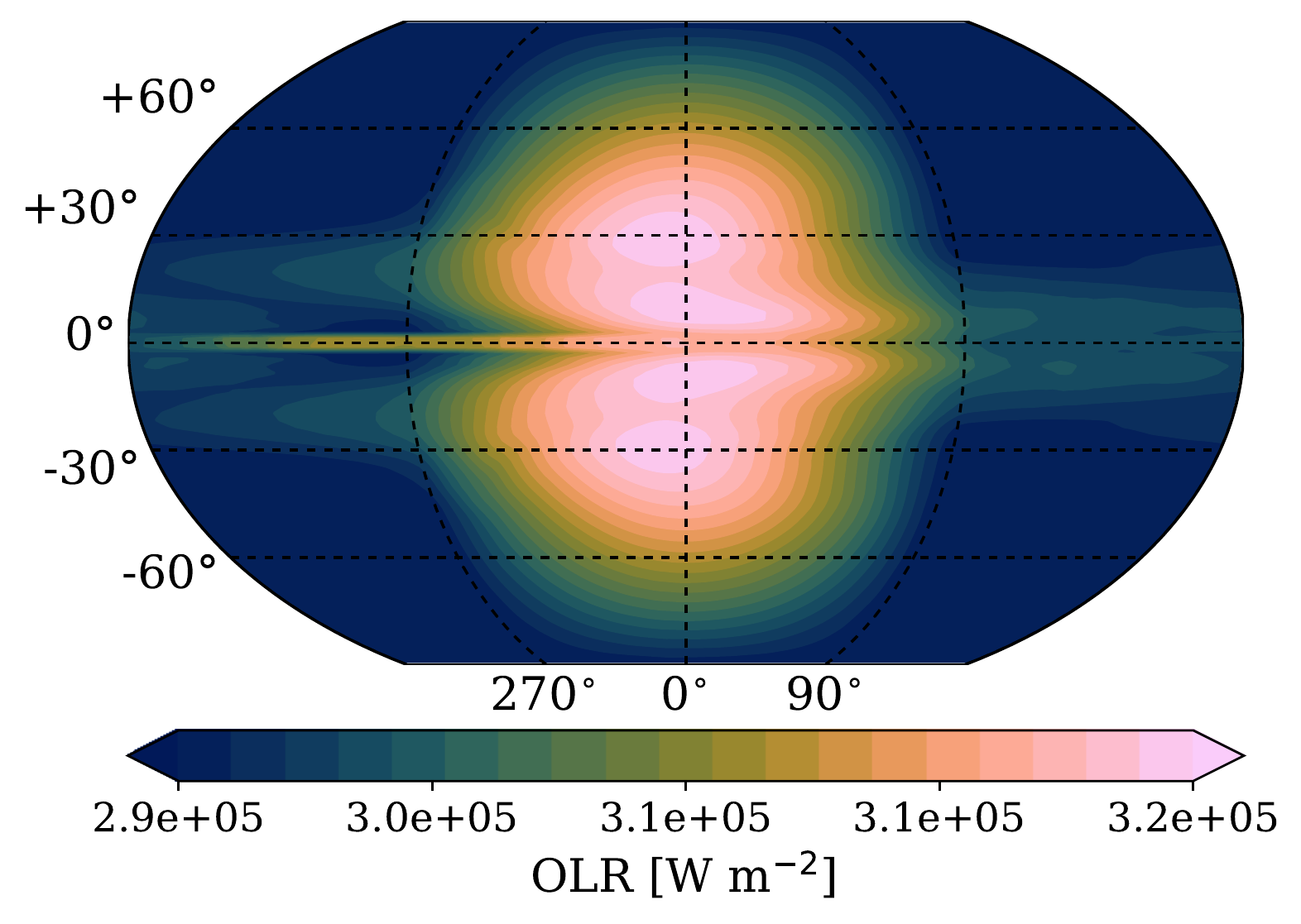}
   \caption{SDSS 1411B GCM results from Exo-FMS using the banded grey opacity scheme.
   Top left: Vertical T-p profiles at the equatorial regions (solid coloured lines) and polar column (dashed).
   Top right: Zonal mean zonal velocity.
   Bottom left: Temperature map at 1 mbar.
   Bottom right: Outgoing longwave radiation (OLR) map.}
   \label{fig:SDSS1411B_GCM_grey}
\end{figure*}

In figure \ref{fig:SDSS1411B_GCM_grey} we show the results of the SDSS 1411B  GCM using the multi-band grey scheme.
Again, as in for the WD0137B model, the multi-band scheme matches well with the 1D RCE modelling from \citet{Lothringer2020}.
However, we fail to match the gradient of the deep adiabat as well as the WD0137B model and the sharpness of the temperature inversion near 10$^{-3}$ bar, as well as the upper atmosphere isothermal region at very low pressures.

Again the zonal mean zonal velocity plot shows a typical BD-WD rotationally dominated dynamical regime, similar to the WD0137B model.
The shape of the OLR and temperature maps differ slightly from the WD0137B, probably due to the different rotational velocities of each BD.
Interestingly, a double lobed structure is produced in this model near the sub-stellar point.
With the lobes closest to the equator a standing Rossby wave pattern.

\begin{figure*} 
   \centering
   \includegraphics[width=0.49\textwidth]{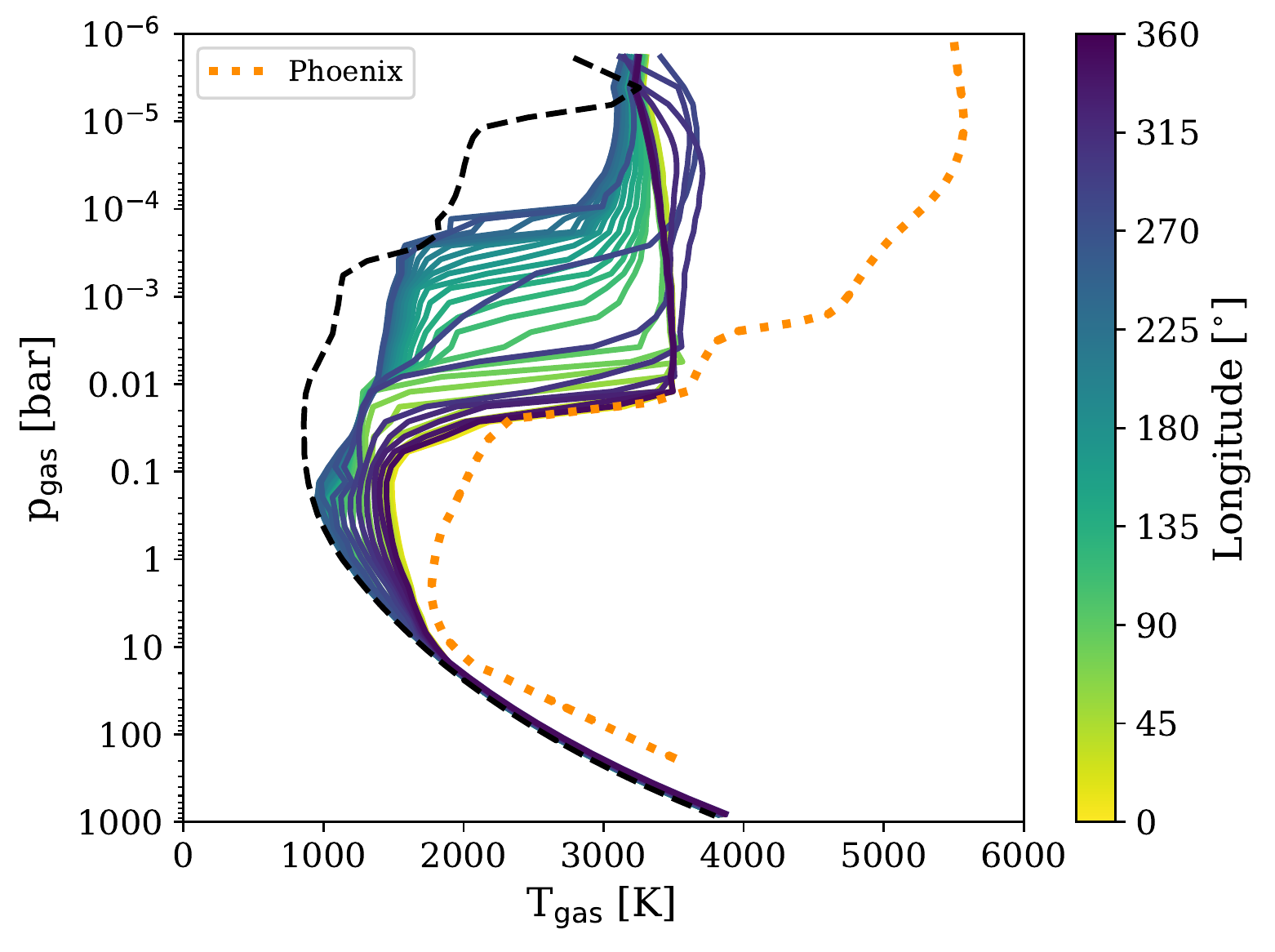}
   \includegraphics[width=0.49\textwidth]{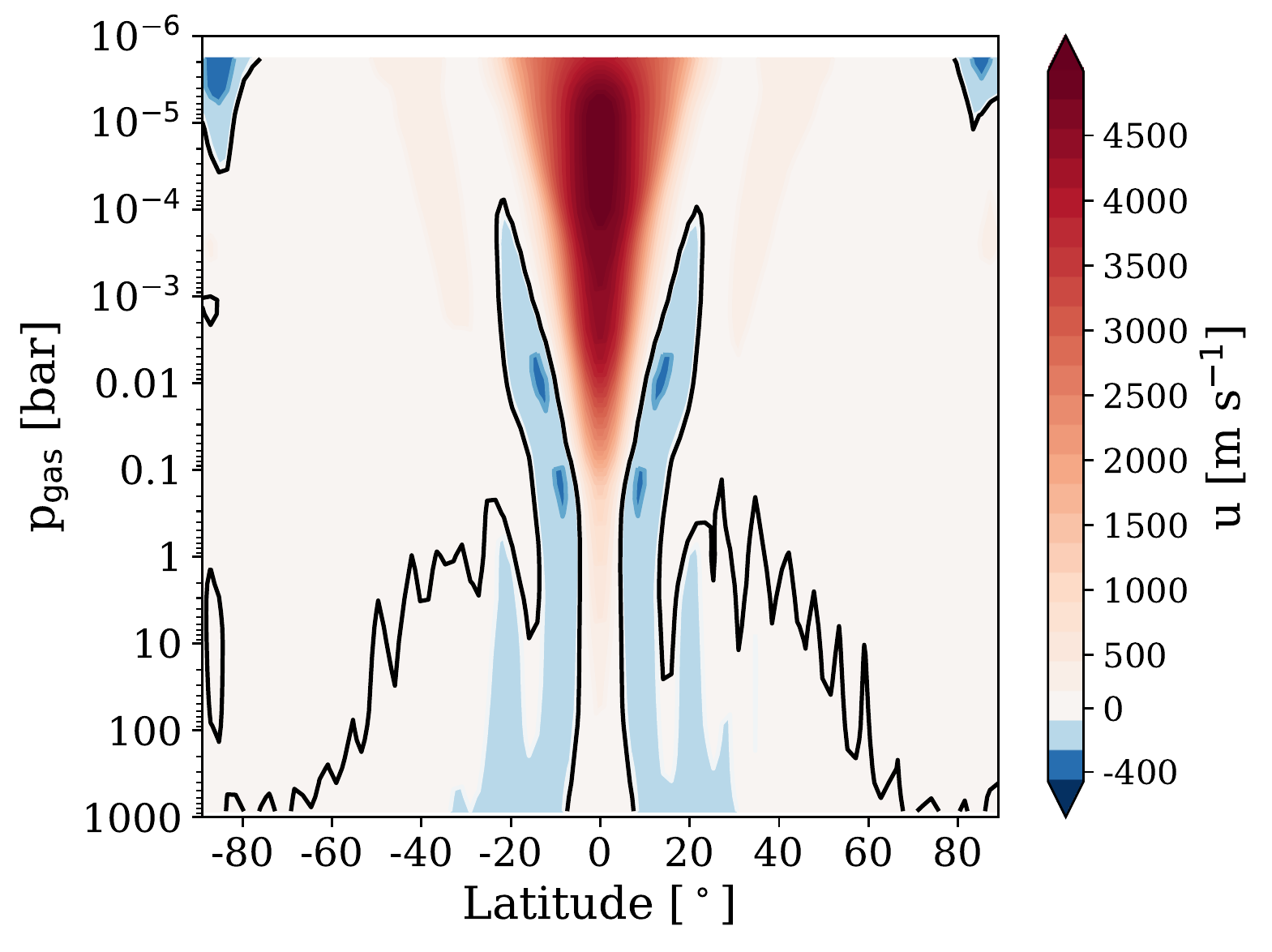}
   \includegraphics[width=0.49\textwidth]{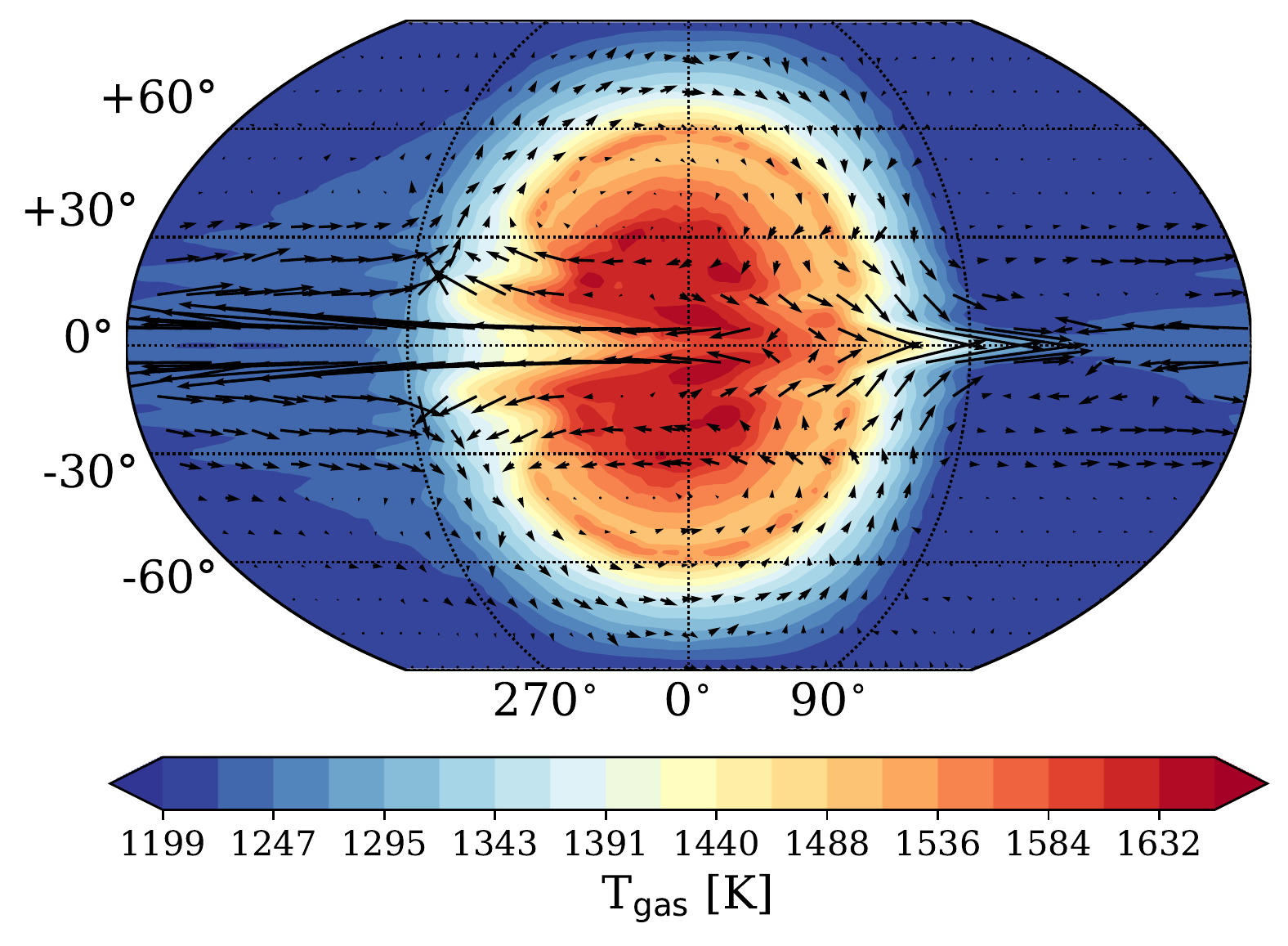}
   \includegraphics[width=0.49\textwidth]{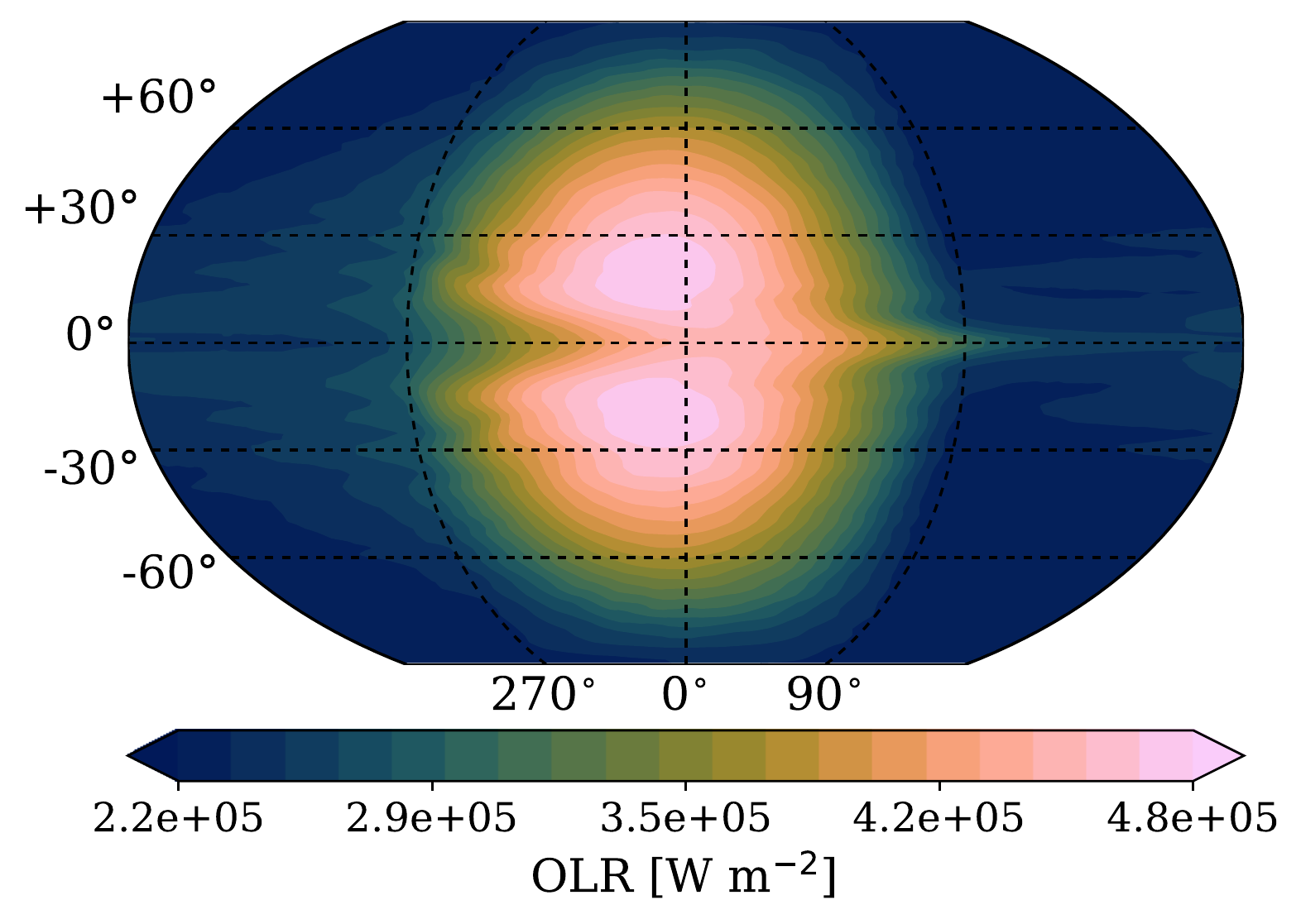}
   \caption{SDSS 1411B GCM results from Exo-FMS using the correlated-k scheme.
   Top left: Vertical T-p profiles at the equatorial regions (solid coloured lines) and polar column (dashed).
   Top right: Zonal mean zonal velocity.
   Bottom left: Temperature map at 1 mbar.
   Bottom right: Outgoing longwave radiation (OLR) map.}
   \label{fig:SDSS1411B_GCM_ck}
\end{figure*}

As with the WD0137B corr-k model, the SDSS 1411B corr-k model (Fig. \ref{fig:SDSS1411B_GCM_ck}) shows sharper T-p structures, but it again fails to reproduce well the very low pressure temperature structures from the 1D model.
The double lobed feature seen in the multi-band grey scheme temperature and OLR maps disappears in the corr-k case.

\subsection{EPIC2122B}

\begin{figure*} 
   \centering
   \includegraphics[width=0.49\textwidth]{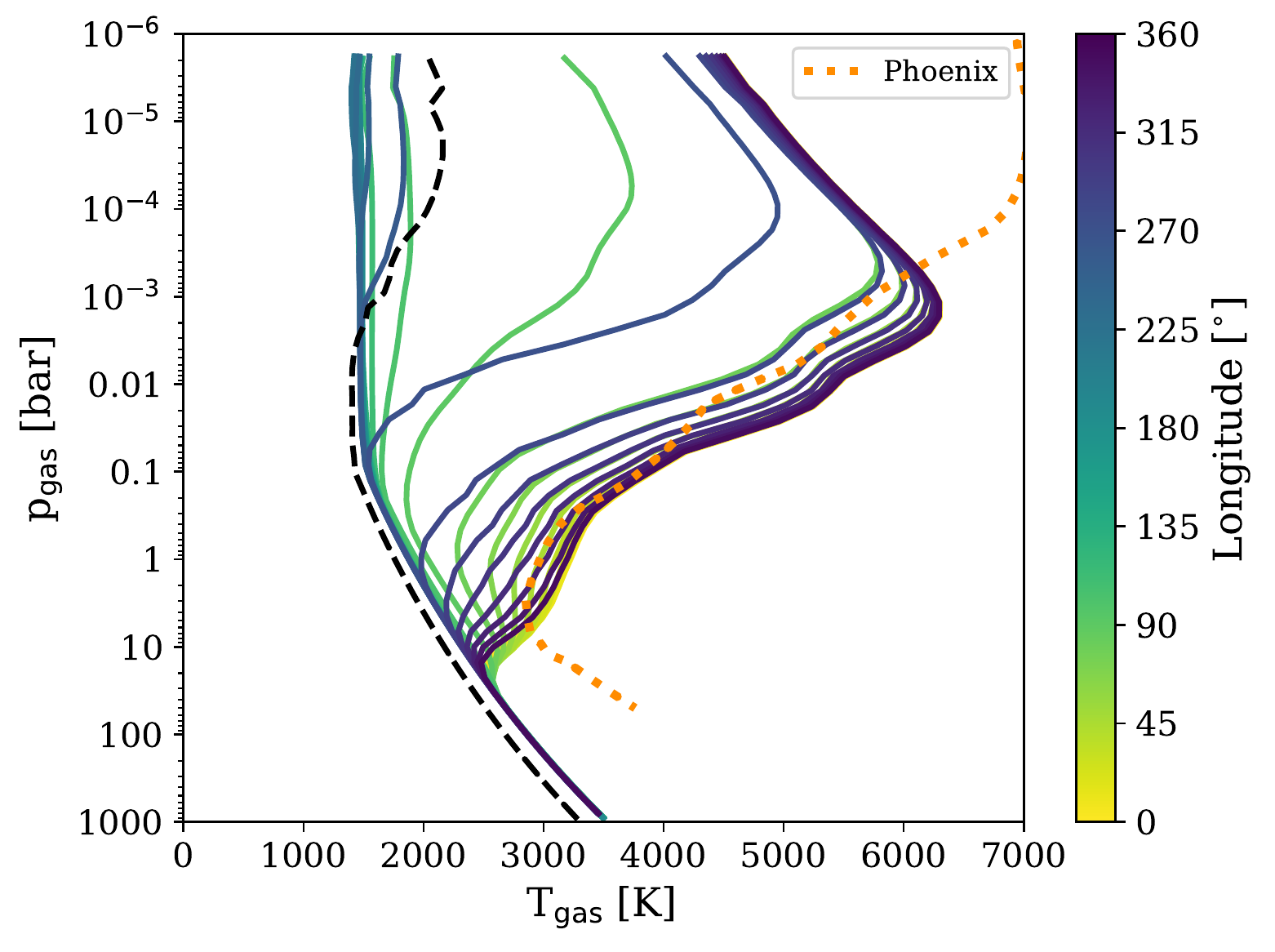}
   \includegraphics[width=0.49\textwidth]{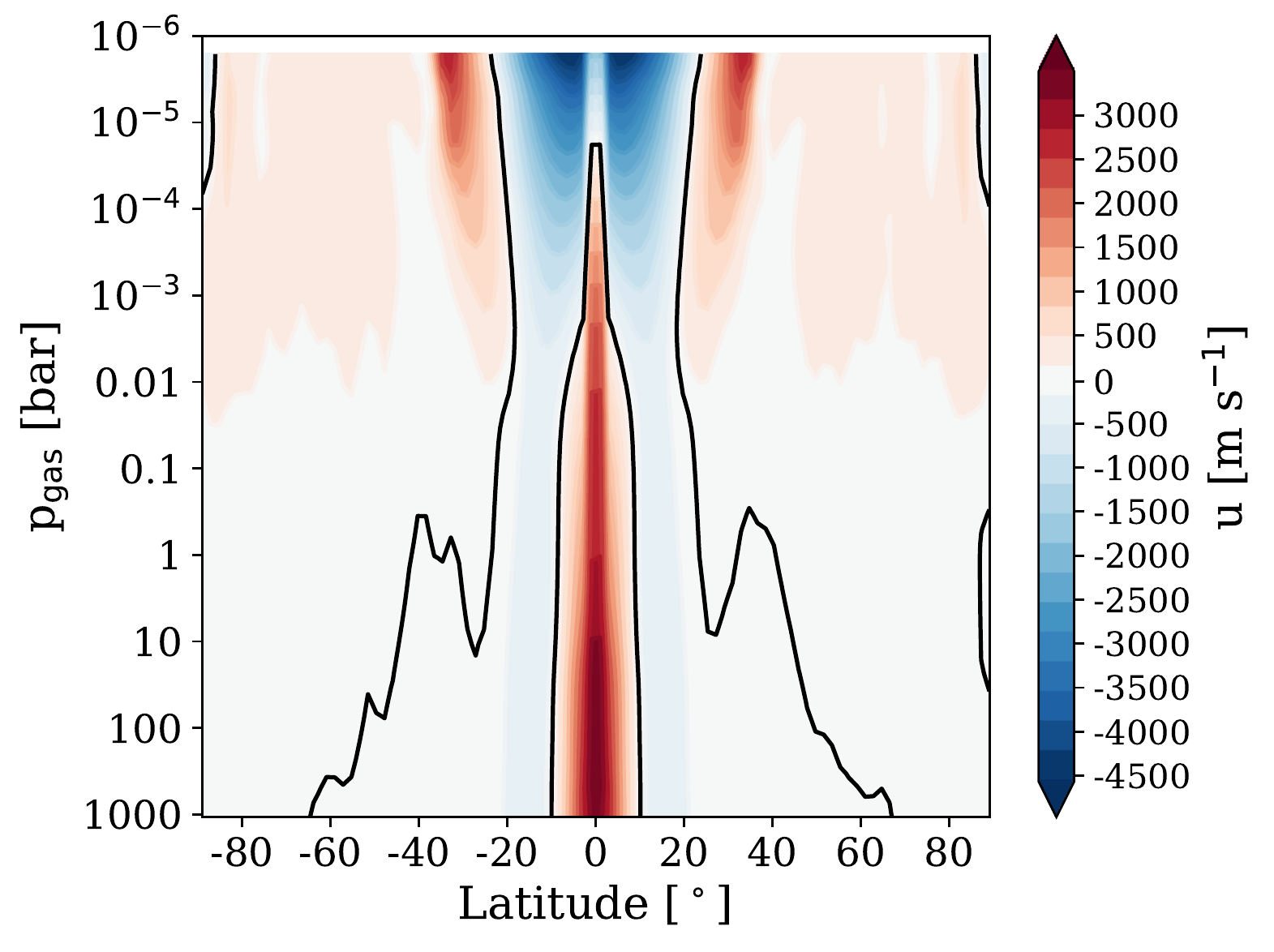}
   \includegraphics[width=0.49\textwidth]{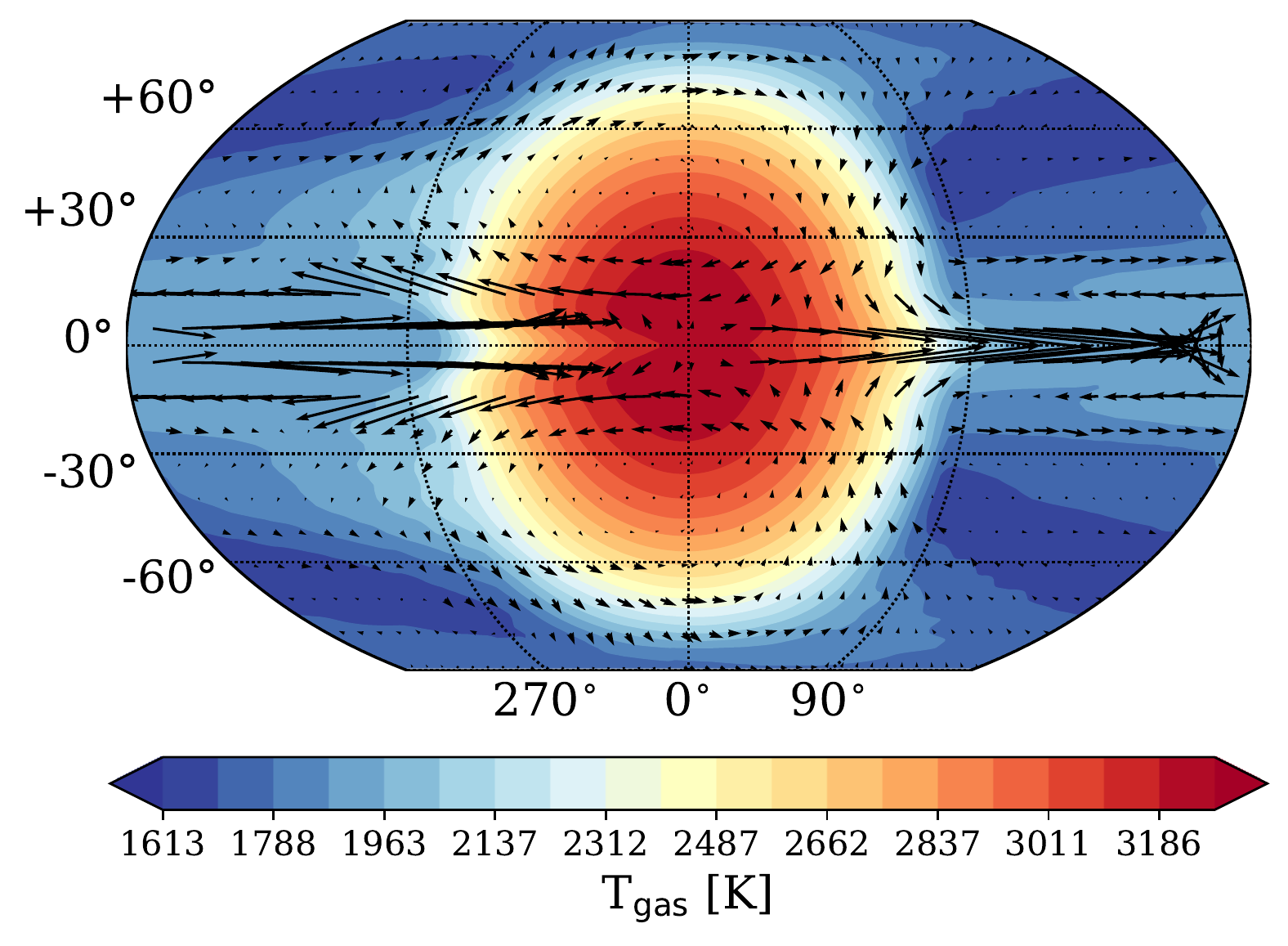}
   \includegraphics[width=0.49\textwidth]{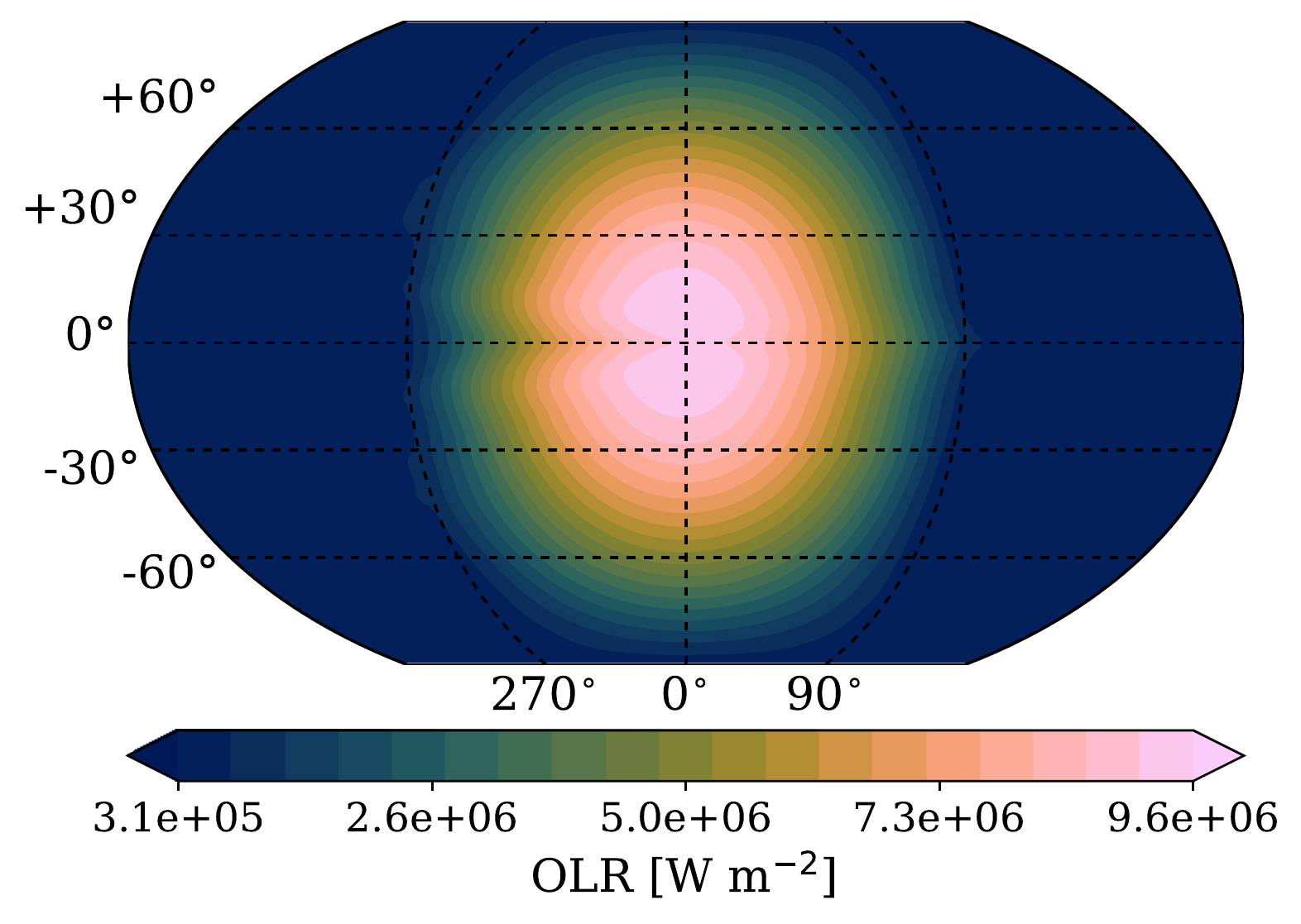}
   \caption{EPIC2122B GCM results from Exo-FMS using the banded grey opacity scheme.
   Top left: Vertical T-p profiles at the equatorial regions (solid coloured lines) and polar column (dashed).
   Top right: Zonal mean zonal velocity.
   Bottom left: Temperature map at 1 mbar.
   Bottom right: Outgoing longwave radiation (OLR) map.}
   \label{fig:EPIC2122B_GCM_grey}
\end{figure*}

In Figure \ref{fig:EPIC2122B_GCM_grey} we show the results of the EPIC2122B GCM using the multi-band grey scheme.
Unlike in the previous cases, our multi-band scheme was not able to capture as well the 1D T-p profiles produced in \citet{Lothringer2020}, especially in the deep regions of the atmosphere.
However, we are able to represent well the gradient of dayside temperature inversion, matching closely with the 1D model until low pressures.

Due to the increased rotation rate of EPIC2122B  the dynamical patterns are slightly different compared to the WD0137B and SDSS 1411B models,
featuring a slightly tighter and deeper equatorial jet.
The shape of the OLR and temperature maps show very tight Rossby lobes near the sub-stellar point, suggesting a highly rotationally dominated dynamical regime.

Our EPIC2122B corr-k model (Fig. \ref{fig:EPIC2122B_GCM_grey}) again does not reproduce the 1D model T-p profile as well, but matches the T-p inversion gradient (however, several hundred Kelvin cooler).
The corr-k model shows tight lobes near the sub-stellar point as in the multi-band grey scheme.

\begin{figure*} 
   \centering
   \includegraphics[width=0.49\textwidth]{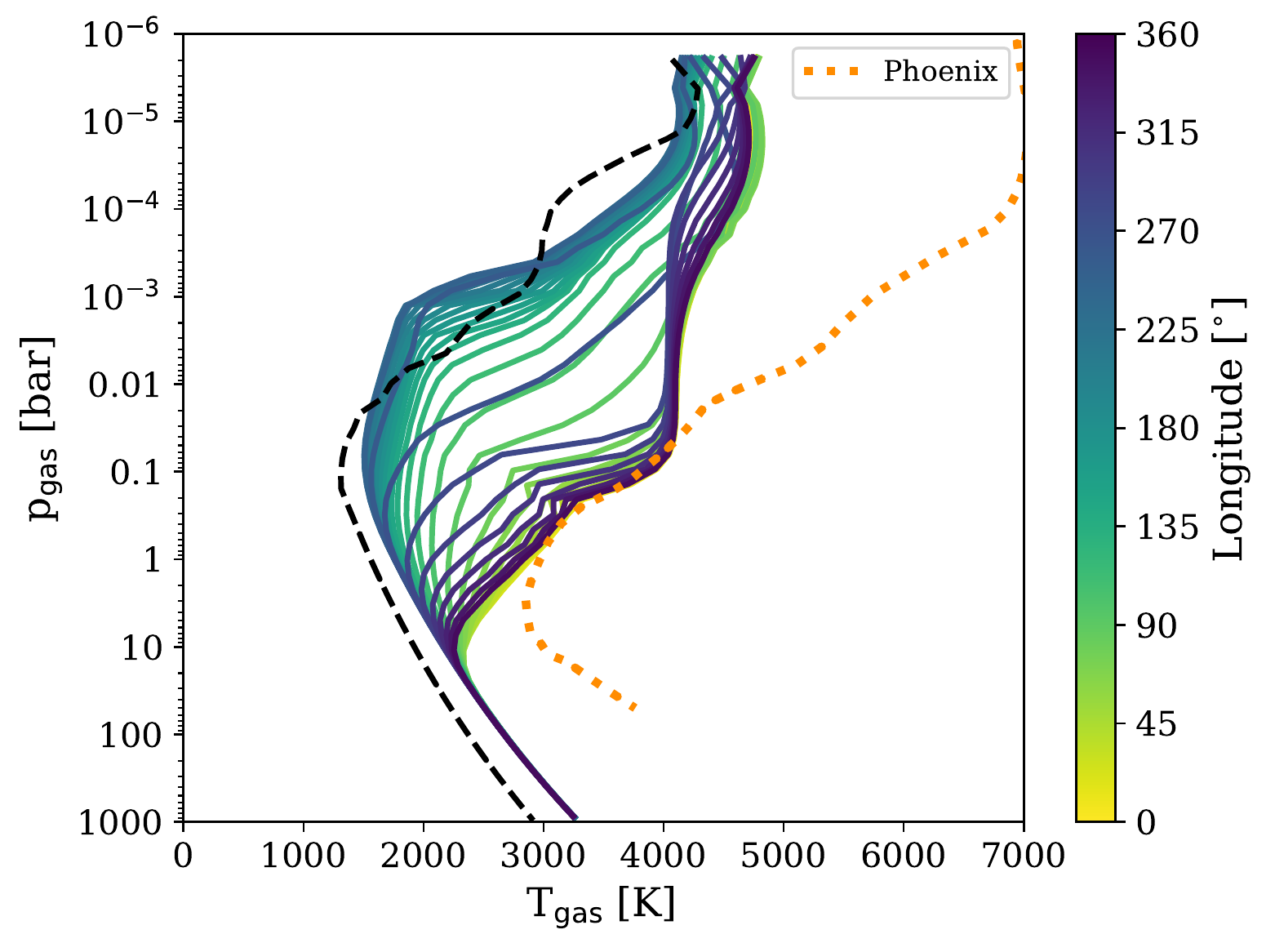}
   \includegraphics[width=0.49\textwidth]{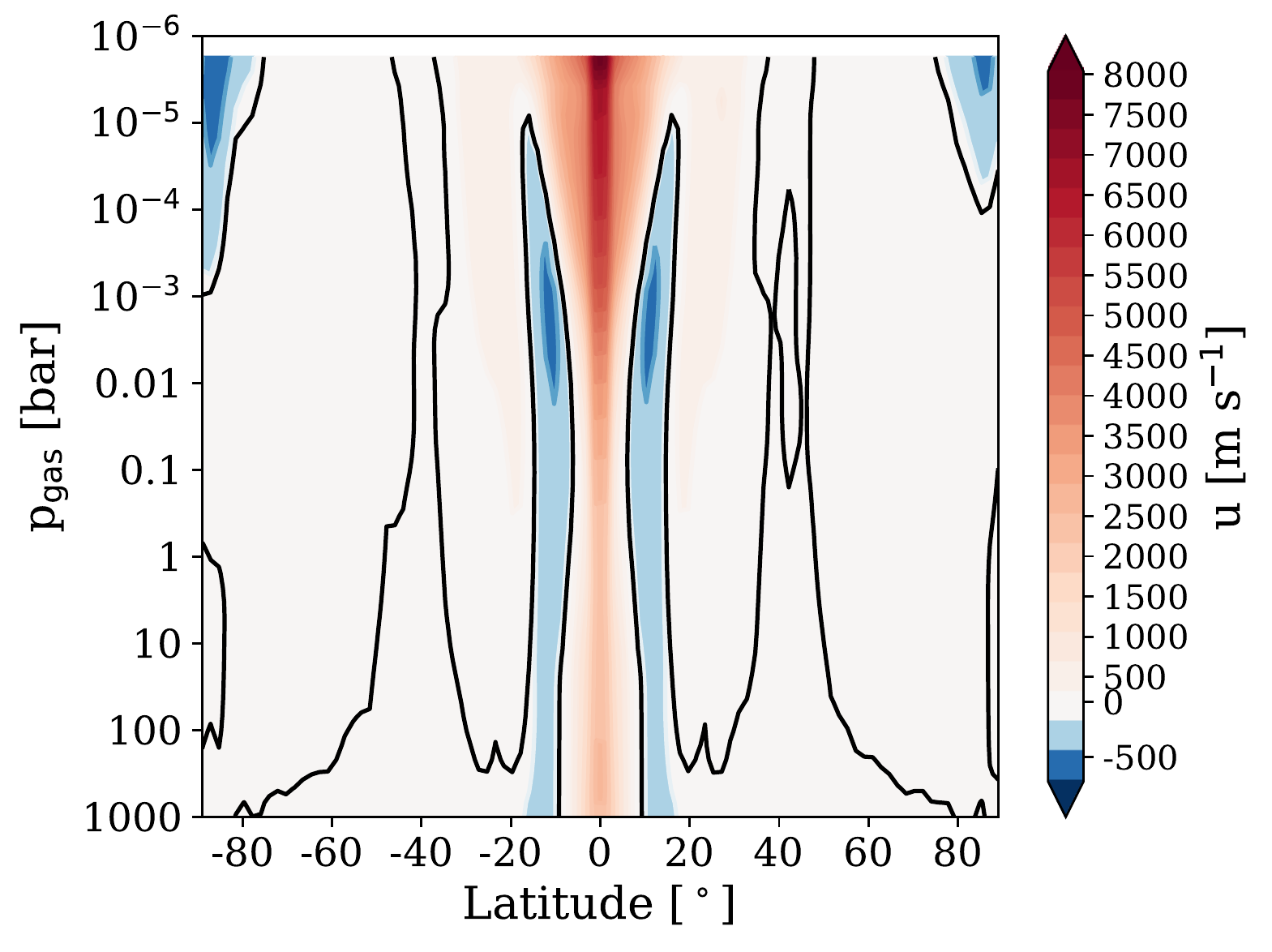}
   \includegraphics[width=0.49\textwidth]{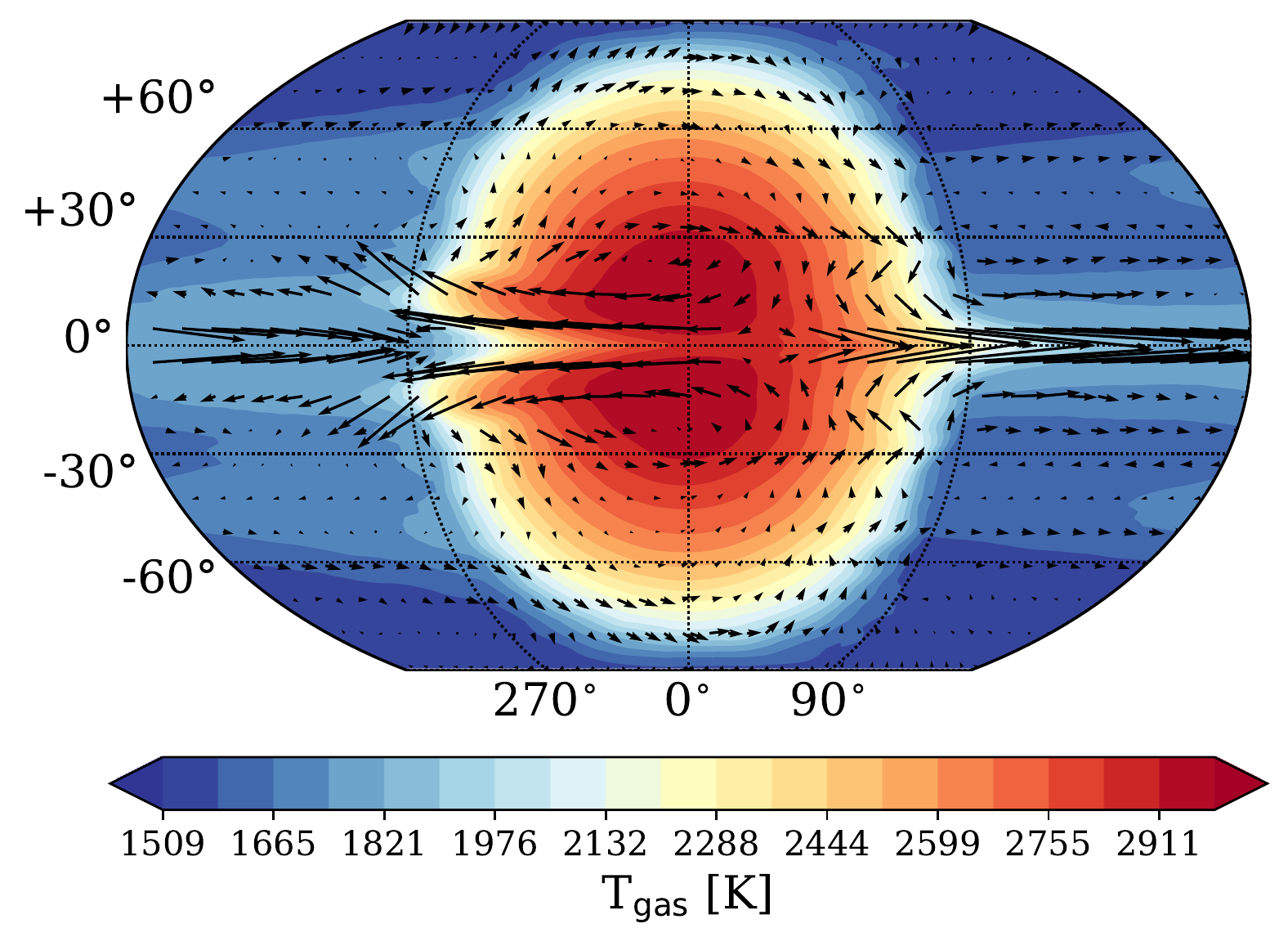}
   \includegraphics[width=0.49\textwidth]{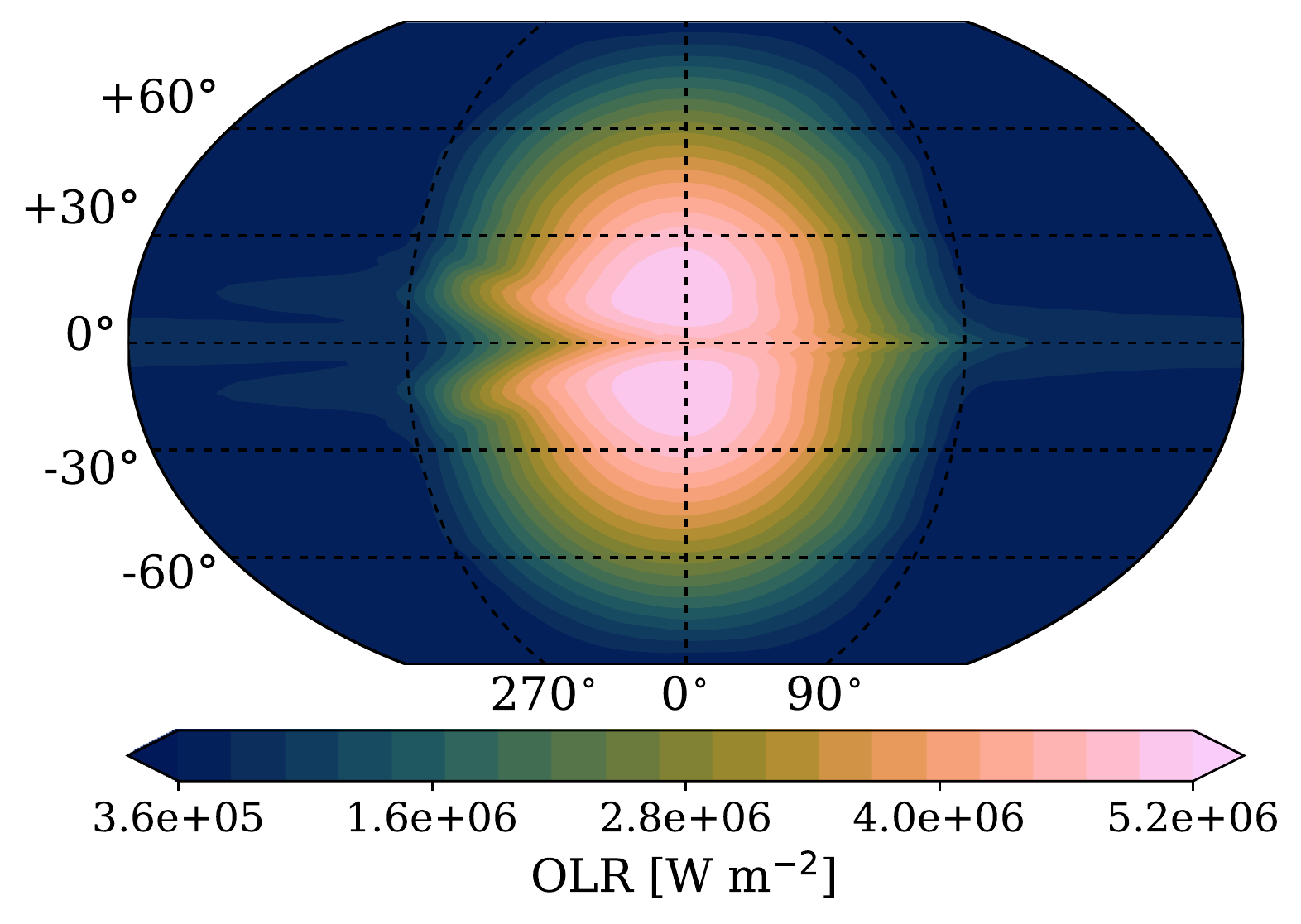}
   \caption{EPIC2122B GCM results from Exo-FMS using the correlated-k scheme.
   Top left: Vertical T-p profiles at the equatorial regions (solid coloured lines) and polar column (dashed).
   Top right: Zonal mean zonal velocity.
   Bottom left: Temperature map at 1 mbar.
   Bottom right: Outgoing longwave radiation (OLR) map.}
   \label{fig:EPIC2122B_GCM_ck}
\end{figure*}

\section{Post-processing results}
\label{sec:PP_res}

In this section, we present the post-processing results of the GCM models which was performed using the 3D gCMCRT model \citep{Lee2021b}.
We chose to focus on the multi-band grey model GCM results, as they better represent the 1D RCE modelling T-p profiles.
Post-processing of the corr-k models showed very similar results to the multi-banded spectra.
We compare to the available photometric data as well as the HST WFC3 data produced by \citet{Lew2022} and \citet{Zhou2022}.

\subsection{WD0137B}

\begin{figure*} 
   \centering
   \includegraphics[width=0.49\textwidth]{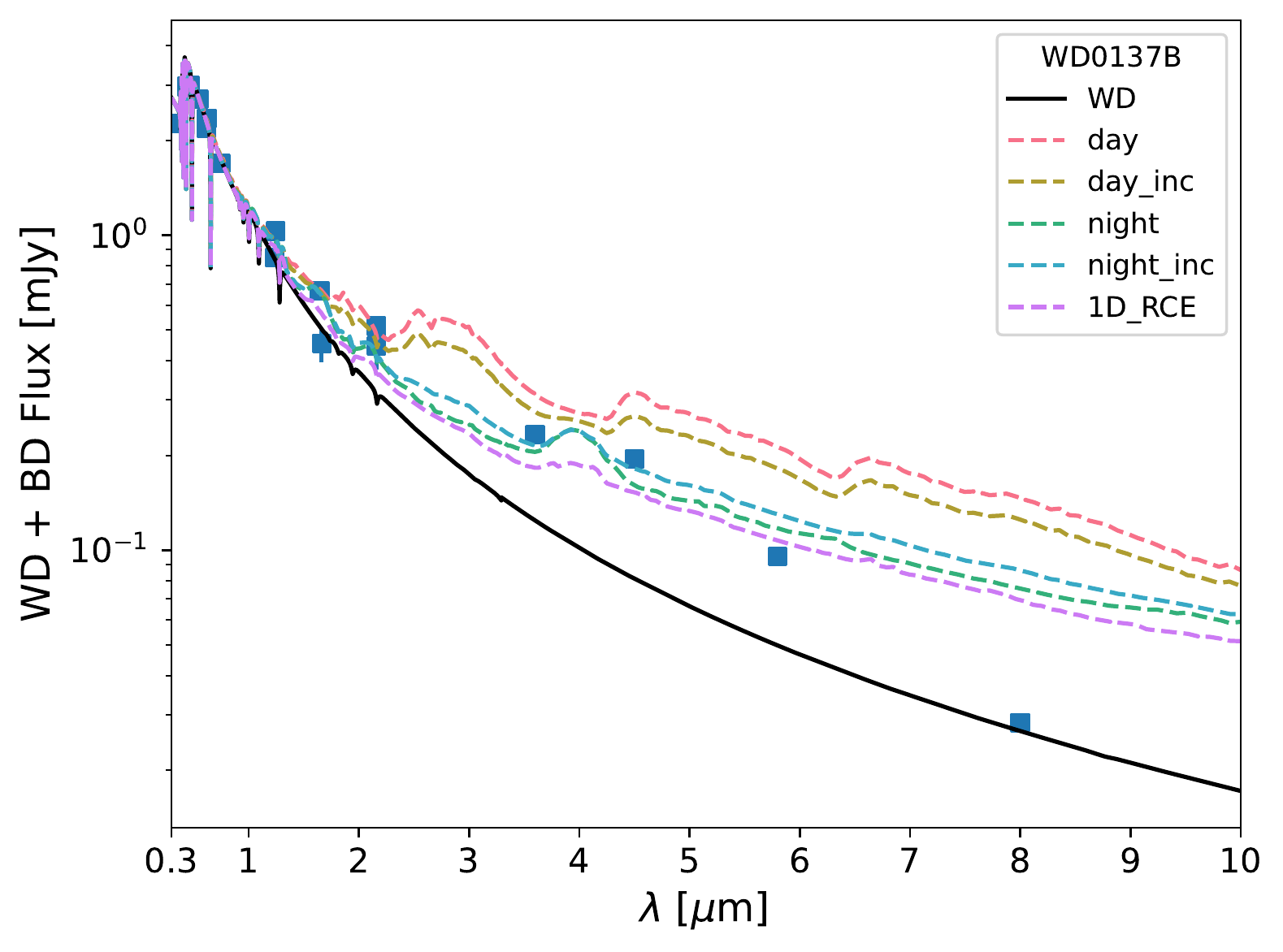}
   \includegraphics[width=0.49\textwidth]{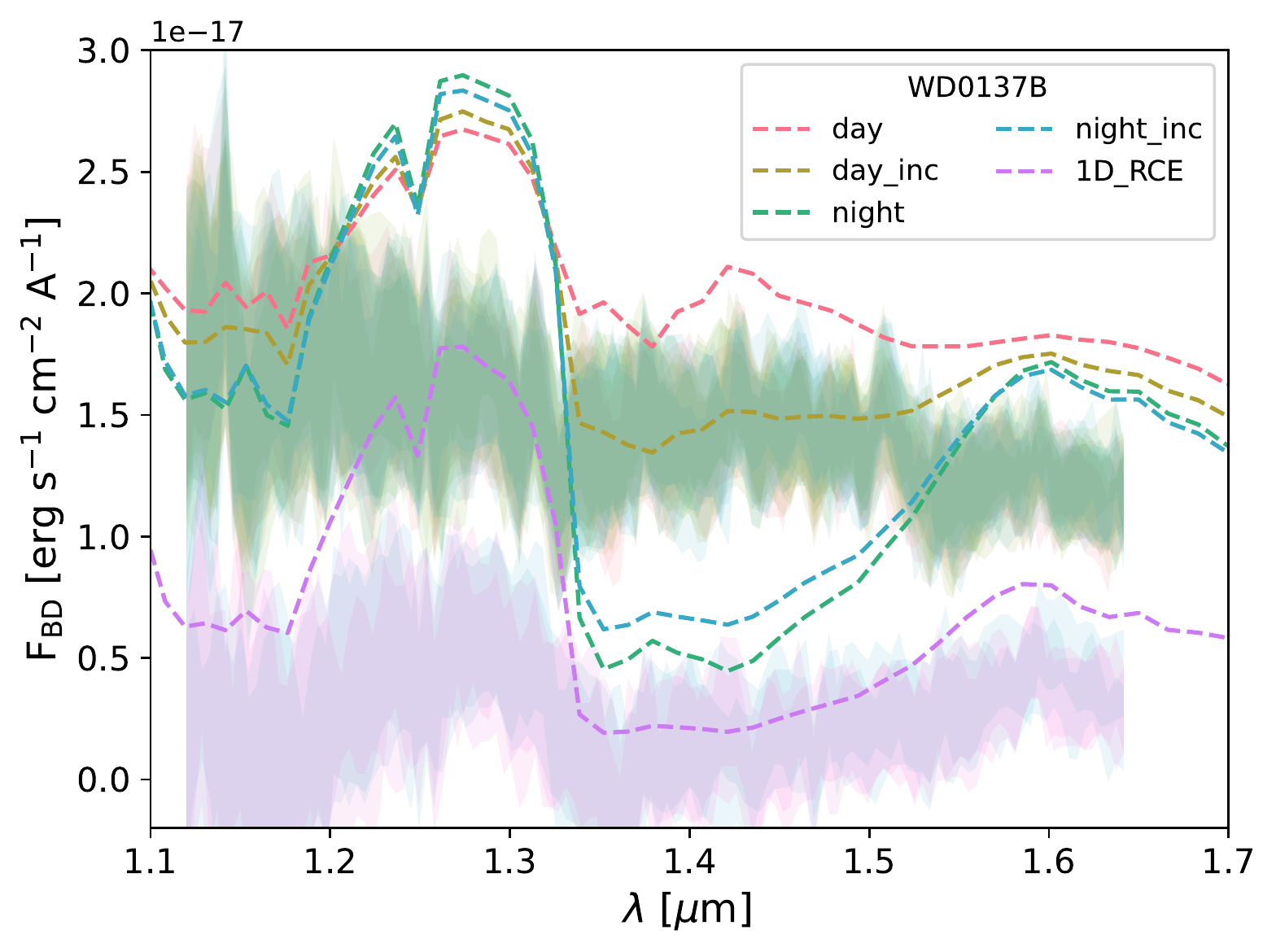}
   \caption{Post-processing results of the WD0137B GCM multi-banded grey model.
   Lines denoted `inc' are the BD spectra at the observed inclination (here 35$^{\circ}$), otherwise the spectra are processed at the equatorial latitude.
   `1D RCE' denotes post-processing of the \citet{Lothringer2020} 1D RCE profile.
   Left: Comparing to the model WD (black line) + BD (coloured dashed lines) total flux to the available photometry data from Gaia EDR3 \citep{Gaia3}, 2MASS \citep{Skrutskie2006} and \citet[][dayside phases]{Casewell2015}.
   Right: Comparing the \citet{Zhou2022} WFC3 dayside phases (green filled region) and nightside phases (purple filled region) spectral data of the BD flux to the model spectra (coloured dashed lines).}
   \label{fig:WD0137B_PP}
\end{figure*}

Figure \ref{fig:WD0137B_PP} shows the post-processing results for the WD0137B multi-band grey GCM model.
Comparing to the photometry of the combined WD+BD flux, the model is able to reproduce well the infrared excess seen in the photometric data.
However, the dayside photometry for the Spitzer bands taken from \citet{Casewell2015} fit better with the nightside spectra of the GCM rather than the dayside spectra, suggesting the dayside is too bright as modelled in the GCM at these wavelengths.

Our comparison with the HST WFC3 data from \citet{Zhou2022} shows that our GCM results do not fit the observed dayside and nightside spectral trends well.
Though our model spectra are at least in the same order of magnitude as the data, on the nightside we significantly overpredict the amplitude of the features as well as their overall flux.
Our dayside spectra also shows large features, not seen in the HST data, with the HST data more consistent with a featureless spectra.
However, our relative difference between the dayside and nightside fluxes at around 1.4$\mu$m seems consistent with the HST data.

\subsection{SDSS 1411B}

\begin{figure*} 
   \centering
   \includegraphics[width=0.49\textwidth]{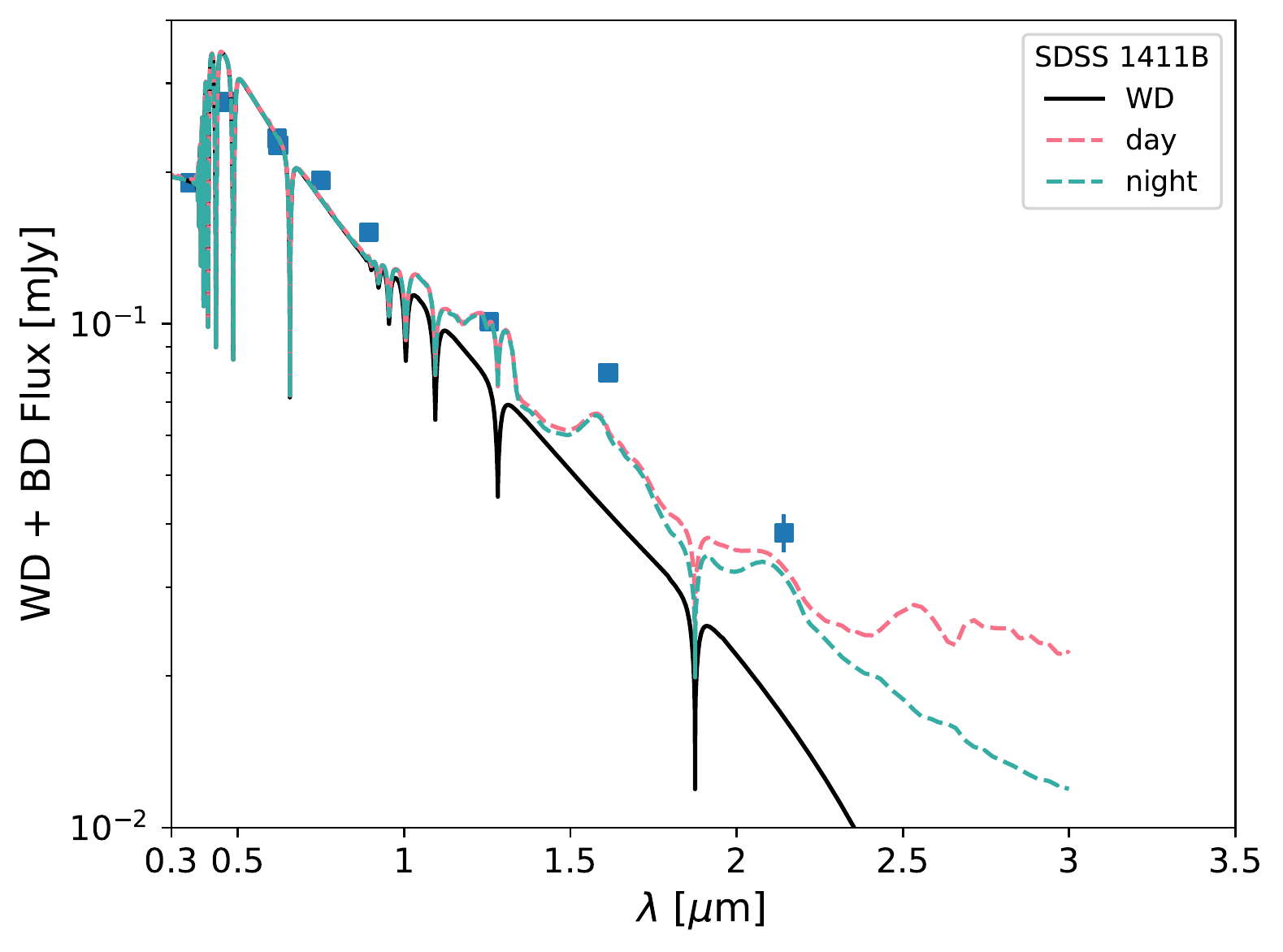}
   \includegraphics[width=0.49\textwidth]{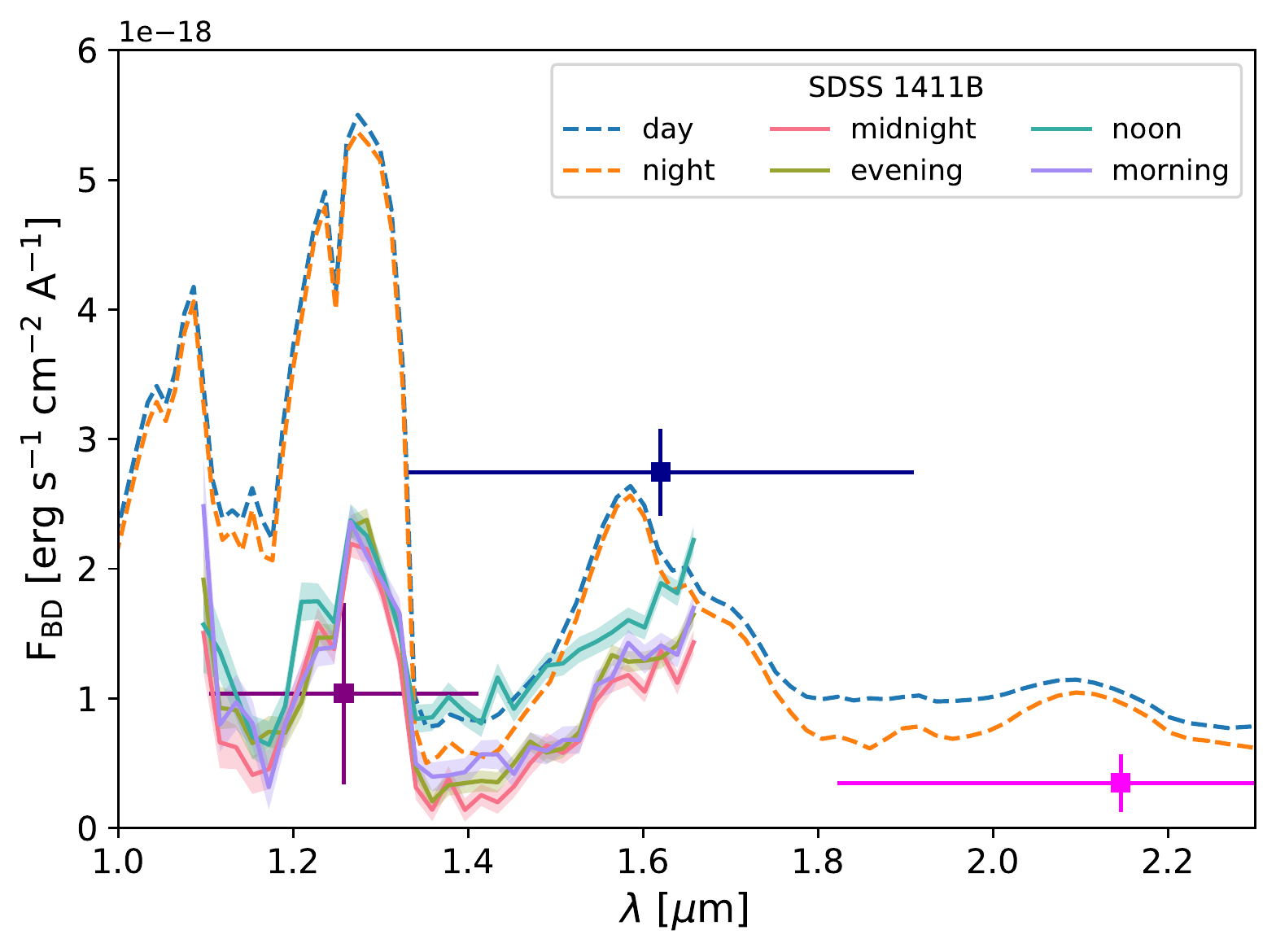}
   \caption{Post-processing results of the SDSS 1411B GCM multi-banded grey model.
   Spectra are processed at the equatorial latitude.
   Left: Comparing to the model WD (black line) + BD (coloured dashed lines) total flux to the available photometry data from Gaia EDR3 \citep{Gaia3}, SDSS \citep{Alam2015} and \citet[][dayside phases]{Casewell2018b}.
   Right: Comparing the \citet{Lew2022} WFC3 morning, evening, noon and midnight phases (filled coloured regions) spectral data of the BD flux to the model spectra (coloured dashed lines), the BD photometry from \citet{Casewell2018b} is also plotted (coloured points).}
   \label{fig:SDSS1411B_PP}
\end{figure*}

Figure \ref{fig:SDSS1411B_PP} shows the post-processing results for the SDSS 1411B multi-band grey GCM model.
We are able to reproduce well the infrared excess trends, fitting well the J band point from \citet{Casewell2018b}.
However, we slightly underpredict the H and K$_{s}$ band photometry from \citet{Casewell2018b}.

Comparing to the HST WFC3 data from \citet{Lew2022} shows that we predict the same spectral trend as the data, but overpredict the emission feature fluxes by about a factor of 2-3.
Interestingly, our model also fits the trend of similar dayside and nightside fluxes seen in the HST data, suggesting the GCM is adequately capturing the day-night temperature contrast in this case.

\subsection{EPIC2122B}

\begin{figure*} 
   \centering
   \includegraphics[width=0.49\textwidth]{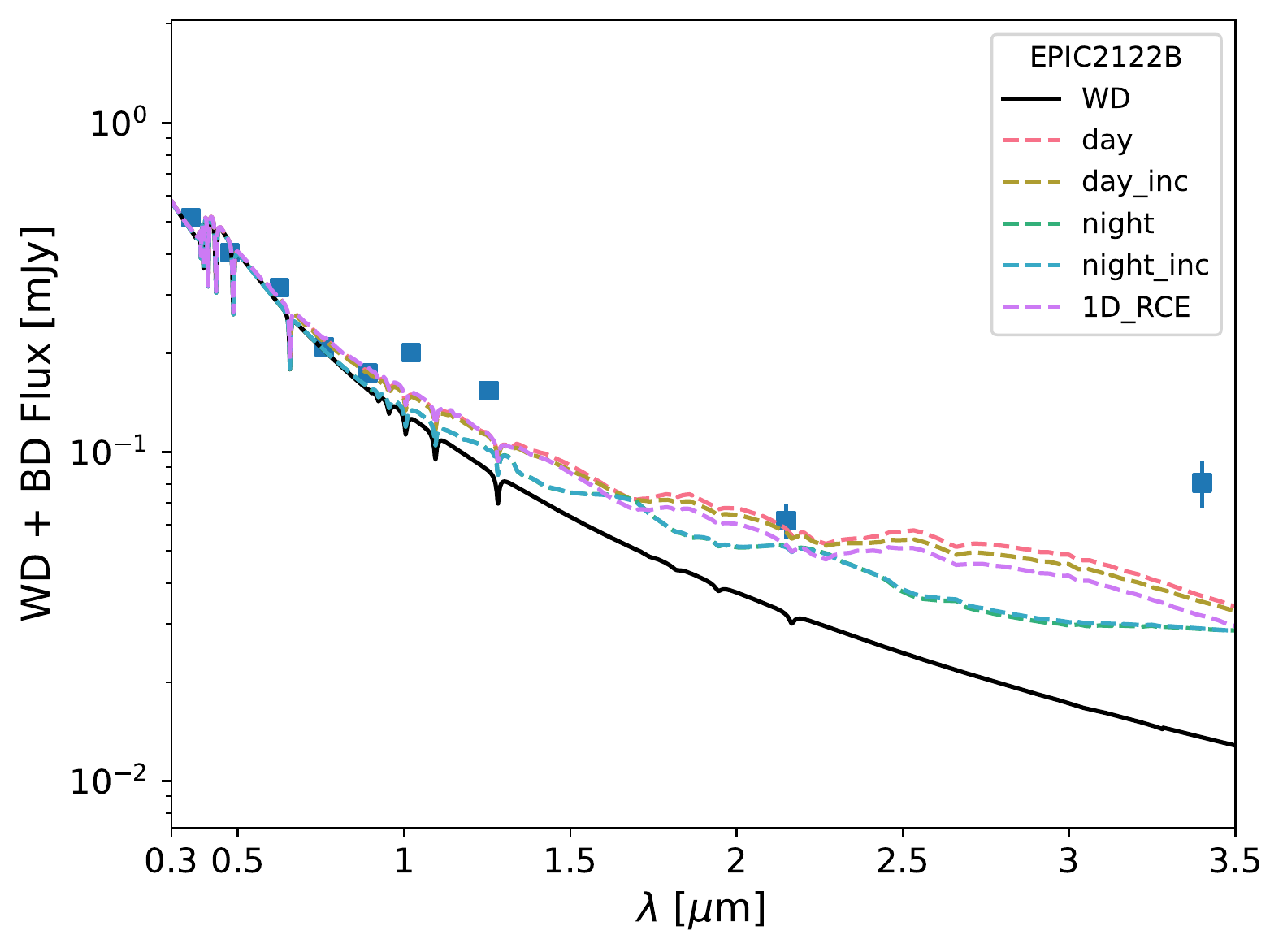}
   \includegraphics[width=0.49\textwidth]{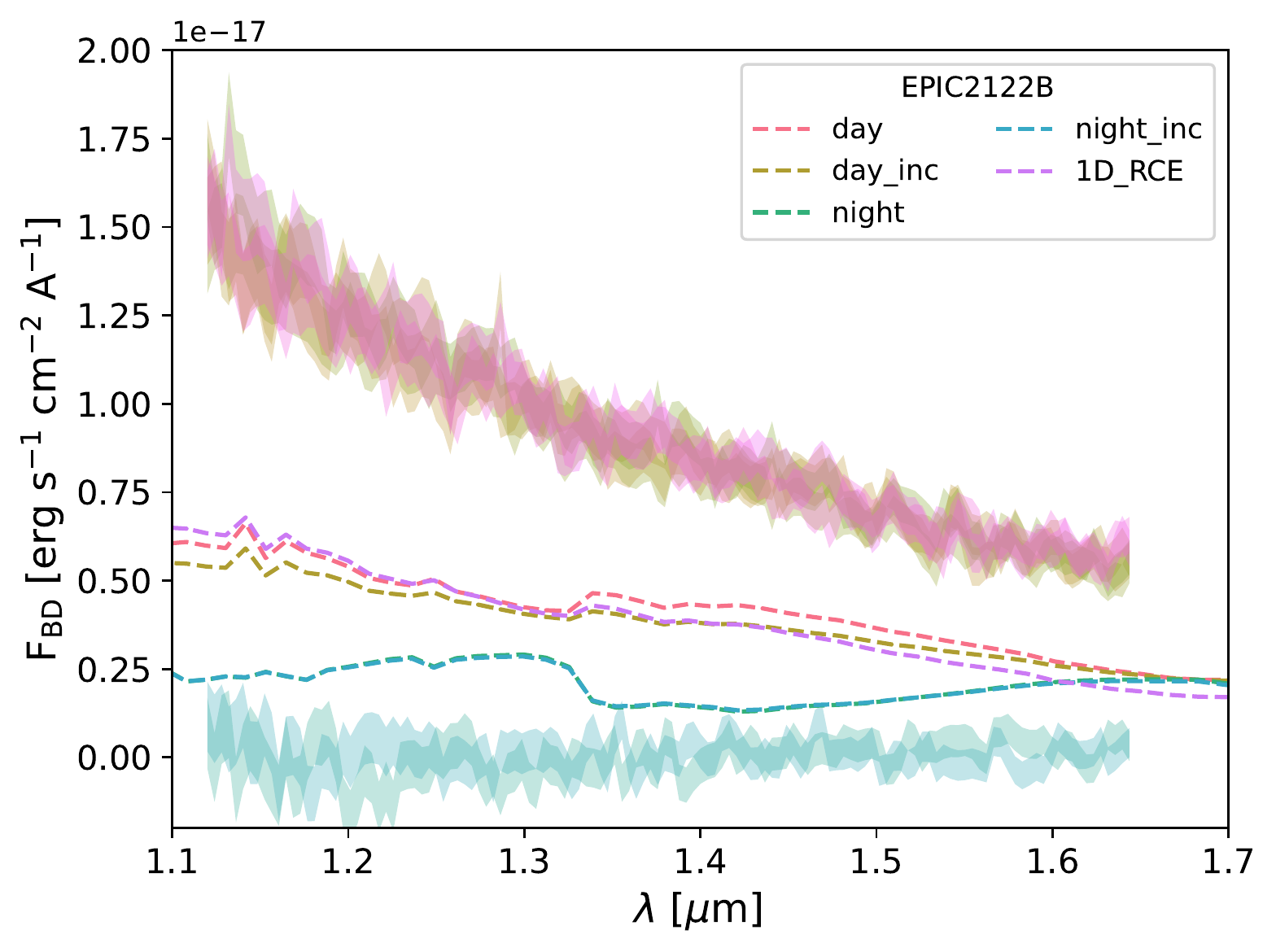}
   \caption{Post-processing results of the EPIC2122B GCM multi-banded grey model.
   Lines denoted `inc' are the BD spectra at the observed inclination (here 56$^{\circ}$), otherwise the spectra are processed at the equatorial latitude.
   `1D RCE' denotes post-processing of the \citet{Lothringer2020} 1D RCE profile.
   Left: Comparing to the model WD (black line) + BD (coloured dashed lines) total flux to the available photometry data from Galex, VST ATLAS, and the VISTA VHS \citep{Casewell2018}.
   Right: Comparing the \citet{Zhou2022} WFC3 dayside phases (yellow/pink filled region) and nightside phases (cyan filled region) spectral data of the BD flux to the model spectra (coloured dashed lines).}
   \label{fig:EPIC2122B_PP}
\end{figure*}

Figure \ref{fig:EPIC2122B_PP} shows the post-processing results for the EPIC2122B multi-band grey GCM model.
Again, the fit to the infrared excess photometric points is overall reasonable, however, with a significant deviation at around 1-1.5 micron wavelengths.

Our fit to the HST WFC3 data from \citet{Zhou2022} is less fitting, with an underprediction of the dayside flux and overprediction of the nightside flux.
This underprediction of the dayside flux is inline with the underprediction of the photometric points across the same wavelength range.
This suggests that our EPIC2122B GCM model is too cold on the dayside, as well as showing too high a day-night energy transport efficiency.
Comparing the GCM T-p profiles to the 1D RCE model suggests that the multi-band scheme failed to deliver adequate energy to the important photospheric regions ($\approx$ 1-10 bar), producing several hundred Kelvin cooler T-p profiles compared to the RCE model.

\section{Discussion}
\label{sec:discussion}

In this study we have used a simplified approach for the opacities in the RT scheme, namely the use of pressure dependent grey IR and UV-Opt band opacity functions and a constant grey opacity flexible `V' band.
This makes these models less physically based than a full spectral RT scheme which plays out in the GCM model results, with the WD0137B and SDSS 1411B models fitting well the 1D RCE profiles from \citet{Lothringer2020}, but the EPIC2122B parameters providing a worse fit when coupled with the GCM.
Possibly an improvement would be to fit to 1D RCE models that assume dayside heat redistribution only, increasing the T$_{\rm irr}$ value used in the GCM model compared to assuming a global redistribution efficiency.
Including a comprehensive opacity treatment such as that used in \citet{Lothringer2020} would be computationally prohibitive inside the GCM.
However, our T-p profiles in all three corr-k models fail to reproduce the very low pressure T-p profiles produced by the 1D models.
This suggests that we are missing important low pressure absorbers in the corr-k models.

Our current correlated-k scheme does not extend into the FUV beyond 0.2$\mu$m, where absorption by H$_{2}$ and CO would occur at high altitude \citep{Sharp2007}.
With H$_{2}$ likely to be thermally and photochemically decomposed into H these regions, we suggest it likely the main missing molecular absorber in this region is CO.
Atomic H and atomic metals can also absorb at high altitudes, making them also candidates for helping maintain a low-pressure temperature inversion \citep{Lothringer2020, Lothringer2020b}.
These are the likely the species required to be included in the k-tables for more accurate spectral modelling of these objects.
Including FUV heating would further reduce the radiative-timescale (and potentially dynamical timescale as well).
This reduction may then make the 3D simulations less stable and computationally feasible.
However, our future models will attempt to consider the effects of FUV heating on the very low pressure atmosphere.

We also do not include the effect of chemical heating and cooling due to the dissociation and recombination of photochemical products.
This may significantly alter the temperature profiles in the upper atmospheres of these brown dwarfs.
The photochemical products may also induce haze formation in some parts of the atmosphere \citep[e.g.][]{Morley2013}.
These hazes may then be dynamically transported across the atmosphere \citep{Steinrueck2021}.
However, our modelled BD systems may be too warm for photochemical hazes to be thermally stable but cooler WD-BD or planetary systems may be able to host hazes in their atmospheres.
Our models also do not include a condensation cloud formation or cloud feedback scheme, known to be important components on the nightside of UHJ and HJ atmospheres \citep[e.g.][]{Helling2021, Parmentier2021}.
Such cloud formation may be important in the polar and nightside regions of the brown dwarfs, reducing the outgoing flux from these regions.
This was also discussed in \citet{Lee2020}, where the nightside fluxes were also too large in comparison to the observational data.

We do not include the effect of H$_{2}$ and H dissociation and recombination on the atmospheric dynamics as performed for UHJs in \citet{Tan2019}.
This effect may be doubly important for WD-BD objects due to the added possibility of dissociation of H$_{2}$ from photochemical effects as well as thermal dissociation.
From our kinetic-chemical post-processing in \citet{Lee2020}, the upper atmosphere of these brown dwarf objects is highly ionised from thermochemical and photochemical processes.
The H$_{2}$ dissociation fraction may therefore approach 100\%.

From our GCM results and post-processing, it is clear in most cases the GCM results are predicting a too low a day-night temperature contrast compared to the observations.
A traditional explanation in the HJ literature suggests that the day-night energy transport is too efficient in the GCMs \citep[e.g.][]{Komacek2017}, driven by a balance between the dynamical and radiative timescales, in addition to any atmospheric drag mechanisms present in the atmosphere.
However, the previous studies of these objects \citep{Tan2020,Lee2020,Zhou2022} suggest that the dayside and nightside T-p profiles are very close to radiative-equilibrium, as well as the rotationally dominated dynamical regime of these objects, naturally producing poor day-night energy transport efficiencies.

Due to the highly ionised nature of these BD atmospheres \citep{Lee2020}, we suggest that magnetic drag effects on the atmosphere is a vital component in setting the dynamical and thermal properties of these BDs.
This would be most apparent in the upper atmosphere, where the ionisation fractions can be extremely high \citep{Lee2020}.
\citet{Beltz2022} show that even small magnetic drag can have a large effect on the atmospheric dynamical properties of HJ atmospheres, suggesting that similar behaviour to that seen in \citet{Beltz2022} is probably occurring in our WD-BD cases.

Clouds may also provide an important effect in setting the nightside and dayside fluxes of the BDs.
Since we overpredict the nightside fluxes (the same as in \citet{Lee2020}) and the T-p profiles of these objects overlap with high temperature condensates such as CaTiO$_{3}$, Al$_{2}$O$_{3}$, Fe and Mg$_{2}$SiO$_{4}$ \citep[e.g.][]{Morley2012, Wakeford2017a}, especially in the photospheric regions from 1-10 bar below the temperature inversion areas.
The formation of clouds would act the reduce the outgoing flux, as as mute spectral features in emission.
SSDSS 1411B is also possibly cool enough on the dayside to also host cloud particulates, potentially also reducing the outgoing flux on the dayside as well.
Due to the high gravity and small vertical winds in these objects \citep{Lee2020} we foresee that such clouds would have to consist of mostly sub-micron particles to remain suspended for long timescales in the atmosphere \citep[e.g.][]{Marley2000, Marley2002}.

\section{Summary \& Conclusions}
\label{sec:conclusions}

Improving on the initial modelling performed in \citet{Lee2020}, using the Exo-FMS GCM model, we investigated the 3D atmospheric structures of the highly irradiated brown dwarfs, WD0137B, SDSS 1411B and EPIC2122B, all three of which orbit close to their host white dwarf stars.
We couple a multi-band grey RT scheme, which was able to adequately capture the gross UV-Optical absorption from the host white dwarf in line with previous 1D RCE modelling.
We then also used a spectral correlated-k scheme with a high temperature opacity table in an attempt to more self-consistently model the RT.
Our results suggest that further development of the multi-band grey scheme looks like a promising way to explore the extreme radiation present in these systems in a simple and effective way.

Our GCM results show steeply temperature inverted atmospheres due to the strong UV irradiation being absorbed at low pressures.
However, our corr-k scheme was unable to reproduce well the 1D RCE T-p profiles at very low pressure, possibly from missing FUV opacities such as CO in the RT scheme.
The GCM results show typical rotationally dominated dynamical patterns, with thin equatorial jets and multiple jets at latitude.
Hot Rossby lobes are pushed towards the equator, bounding the sub-stellar point region.
Our post-processing results suggest that the GCM models are producing too low a day-night contrast for the BD atmospheres, suggesting the cloud formation near the IR photospheric regions on the nightside (and sometimes dayside if cool enough) may play an important role in shaping the outgoing flux for thee objects.
In addition, we suggest that atmospheric drag effects, such as magnetic drag, will be an important consideration for future modelling efforts.

Overall, our current project reveals further additional complexity when modelling WD-BD system in 3D and informs future modelling efforts on important physical processes to consider when modelling these objects.
WD-BD systems continue to push contemporary modelling capabilities to the limit and provide a unique perspective on the interplay between the atmospheric dynamics and UV radiation in fast rotating objects.

\clearpage

\section*{Data and code availability}

The radiative-transfer schemes are available on GitHub: \url{https://github.com/ELeeAstro}.
The Exo-FMS GCM model and post-processing output is available from the lead author upon request.

\section*{Acknowledgements}
E.K.H. Lee is supported by the SNSF Ambizione Fellowship grant (\#193448).
S.L.C. acknowledges the support of an STFC Ernest Rutherford Fellowship.
Development of the 1D models was supported by program HST-AR-16142.
Support for Program HST-AR-16142 were provided by NASA through a grant from the Space Telescope Science Institute, which is operated by the Association of Universities for Research in Astronomy, Incorporated, under NASA contract NAS5-26555.
Plots were produced using the community open-source Python packages Matplotlib \citep{Hunter2007}, SciPy \citep{Jones2001}, and AstroPy \citep{Astropy2018}.
The HPC support staff at AOPP, University of Oxford and University of Bern are highly acknowledged.




\bibliographystyle{mnras}
\bibliography{bib2} 

\begin{thebibliography}{}
\makeatletter
\relax
\def\mn@urlcharsother{\let\do\@makeother \do\$\do\&\do\#\do\^\do\_\do\%\do\~}
\def\mn@doi{\begingroup\mn@urlcharsother \@ifnextchar [ {\mn@doi@}
  {\mn@doi@[]}}
\def\mn@doi@[#1]#2{\def\@tempa{#1}\ifx\@tempa\@empty \href
  {http://dx.doi.org/#2} {doi:#2}\else \href {http://dx.doi.org/#2} {#1}\fi
  \endgroup}
\def\mn@eprint#1#2{\mn@eprint@#1:#2::\@nil}
\def\mn@eprint@arXiv#1{\href {http://arxiv.org/abs/#1} {{\tt arXiv:#1}}}
\def\mn@eprint@dblp#1{\href {http://dblp.uni-trier.de/rec/bibtex/#1.xml}
  {dblp:#1}}
\def\mn@eprint@#1:#2:#3:#4\@nil{\def\@tempa {#1}\def\@tempb {#2}\def\@tempc
  {#3}\ifx \@tempc \@empty \let \@tempc \@tempb \let \@tempb \@tempa \fi \ifx
  \@tempb \@empty \def\@tempb {arXiv}\fi \@ifundefined
  {mn@eprint@\@tempb}{\@tempb:\@tempc}{\expandafter \expandafter \csname
  mn@eprint@\@tempb\endcsname \expandafter{\@tempc}}}

\bibitem[\protect\citeauthoryear{{Alam} et~al.,}{{Alam}
  et~al.}{2015}]{Alam2015}
{Alam} S.,  et~al., 2015, \mn@doi [\apjs] {10.1088/0067-0049/219/1/12}, \href
  {https://ui.adsabs.harvard.edu/abs/2015ApJS..219...12A} {219, 12}

\bibitem[\protect\citeauthoryear{{Amundsen} et~al.,}{{Amundsen}
  et~al.}{2016}]{Amundsen2016}
{Amundsen} D.~S.,  et~al., 2016, \mn@doi [\aap] {10.1051/0004-6361/201629183},
  \href {http://adsabs.harvard.edu/abs/2016A%26A...595A..36A} {595, A36}

\bibitem[\protect\citeauthoryear{{Amundsen}, {Tremblin}, {Manners}, {Baraffe}
  \& {Mayne}}{{Amundsen} et~al.}{2017}]{Amundsen2017}
{Amundsen} D.~S.,  {Tremblin} P.,  {Manners} J.,  {Baraffe} I.,   {Mayne}
  N.~J.,  2017, \mn@doi [\aap] {10.1051/0004-6361/201629322}, \href
  {https://ui.adsabs.harvard.edu/#abs/2017A&A...598A..97A} {598, A97}

\bibitem[\protect\citeauthoryear{{Asplund}, {Grevesse}, {Sauval}  \&
  {Scott}}{{Asplund} et~al.}{2009}]{Asplund2009}
{Asplund} M.,  {Grevesse} N.,  {Sauval} A.~J.,   {Scott} P.,  2009, \mn@doi
  [\araa] {10.1146/annurev.astro.46.060407.145222}, \href
  {http://adsabs.harvard.edu/abs/2009ARA%26A..47..481A} {47, 481}

\bibitem[\protect\citeauthoryear{{Azzam}, {Tennyson}, {Yurchenko}  \&
  {Naumenko}}{{Azzam} et~al.}{2016}]{Azzam2016}
{Azzam} A. A.~A.,  {Tennyson} J.,  {Yurchenko} S.~N.,   {Naumenko} O.~V.,
  2016, \mn@doi [\mnras] {10.1093/mnras/stw1133}, \href
  {https://ui.adsabs.harvard.edu/abs/2016MNRAS.460.4063A} {460, 4063}

\bibitem[\protect\citeauthoryear{{Barber}, {Strange}, {Hill}, {Polyansky},
  {Mellau}, {Yurchenko}  \& {Tennyson}}{{Barber} et~al.}{2014}]{Barber2014}
{Barber} R.~J.,  {Strange} J.~K.,  {Hill} C.,  {Polyansky} O.~L.,  {Mellau}
  G.~C.,  {Yurchenko} S.~N.,   {Tennyson} J.,  2014, \mn@doi [\mnras]
  {10.1093/mnras/stt2011}, \href
  {https://ui.adsabs.harvard.edu/abs/2014MNRAS.437.1828B} {437, 1828}

\bibitem[\protect\citeauthoryear{Bell}{Bell}{1980}]{Bell1980}
Bell K.~L.,  1980, \mn@doi [Journal of Physics B: Atomic and Molecular Physics]
  {10.1088/0022-3700/13/9/016}, 13, 1859

\bibitem[\protect\citeauthoryear{{Bell}, {Berrington}  \& {Croskery}}{{Bell}
  et~al.}{1982}]{Bell1982}
{Bell} K.~L.,  {Berrington} K.~A.,   {Croskery} J.~P.,  1982, \mn@doi [Journal
  of Physics B Atomic Molecular Physics] {10.1088/0022-3700/15/6/022}, \href
  {https://ui.adsabs.harvard.edu/abs/1982JPhB...15..977B} {15, 977}

\bibitem[\protect\citeauthoryear{{Beltz}, {Rauscher}, {Roman}  \&
  {Guilliat}}{{Beltz} et~al.}{2022}]{Beltz2022}
{Beltz} H.,  {Rauscher} E.,  {Roman} M.~T.,   {Guilliat} A.,  2022, \mn@doi
  [\aj] {10.3847/1538-3881/ac3746}, \href
  {https://ui.adsabs.harvard.edu/abs/2022AJ....163...35B} {163, 35}

\bibitem[\protect\citeauthoryear{Bernath}{Bernath}{2020}]{Bernath2020}
Bernath P.~F.,  2020, \mn@doi [Journal of Quantitative Spectroscopy and
  Radiative Transfer] {https://doi.org/10.1016/j.jqsrt.2019.106687}, 240,
  106687

\bibitem[\protect\citeauthoryear{{Beuermann} et~al.,}{{Beuermann}
  et~al.}{2013}]{Beuermann2013}
{Beuermann} K.,  et~al., 2013, \mn@doi [\aap] {10.1051/0004-6361/201322241},
  \href {https://ui.adsabs.harvard.edu/abs/2013A&A...558A..96B} {558, A96}

\bibitem[\protect\citeauthoryear{{Burleigh}, {Hogan}, {Dobbie}, {Napiwotzki}
  \& {Maxted}}{{Burleigh} et~al.}{2006}]{Burleigh2006}
{Burleigh} M.~R.,  {Hogan} E.,  {Dobbie} P.~D.,  {Napiwotzki} R.,   {Maxted}
  P.~F.~L.,  2006, \mn@doi [\mnras] {10.1111/j.1745-3933.2006.00242.x}, \href
  {https://ui.adsabs.harvard.edu/abs/2006MNRAS.373L..55B} {373, L55}

\bibitem[\protect\citeauthoryear{{Carone} et~al.,}{{Carone}
  et~al.}{2020}]{Carone2020}
{Carone} L.,  et~al., 2020, \mn@doi [\mnras] {10.1093/mnras/staa1733}, \href
  {https://ui.adsabs.harvard.edu/abs/2020MNRAS.496.3582C} {496, 3582}

\bibitem[\protect\citeauthoryear{{Casewell} et~al.,}{{Casewell}
  et~al.}{2015}]{Casewell2015}
{Casewell} S.~L.,  et~al., 2015, \mn@doi [\mnras] {10.1093/mnras/stu2721},
  \href {https://ui.adsabs.harvard.edu/#abs/2015MNRAS.447.3218C} {447, 3218}

\bibitem[\protect\citeauthoryear{{Casewell} et~al.,}{{Casewell}
  et~al.}{2018a}]{Casewell2018}
{Casewell} S.~L.,  et~al., 2018a, \mn@doi [\mnras] {10.1093/mnras/sty245},
  \href {https://ui.adsabs.harvard.edu/#abs/2018MNRAS.476.1405C} {476, 1405}

\bibitem[\protect\citeauthoryear{{Casewell}, {Littlefair}, {Parsons}, {Marsh},
  {Fortney}  \& {Marley}}{{Casewell} et~al.}{2018b}]{Casewell2018b}
{Casewell} S.~L.,  {Littlefair} S.~P.,  {Parsons} S.~G.,  {Marsh} T.~R.,
  {Fortney} J.~J.,   {Marley} M.~S.,  2018b, \mn@doi [\mnras]
  {10.1093/mnras/sty2599}, \href
  {https://ui.adsabs.harvard.edu/abs/2018MNRAS.481.5216C} {481, 5216}

\bibitem[\protect\citeauthoryear{{Chubb}, {Tennyson}  \& {Yurchenko}}{{Chubb}
  et~al.}{2020}]{Chubb2020}
{Chubb} K.~L.,  {Tennyson} J.,   {Yurchenko} S.~N.,  2020, \mn@doi [\mnras]
  {10.1093/mnras/staa229}, \href
  {https://ui.adsabs.harvard.edu/abs/2020MNRAS.493.1531C} {493, 1531}

\bibitem[\protect\citeauthoryear{{Coles}, {Yurchenko}  \& {Tennyson}}{{Coles}
  et~al.}{2019}]{Coles2019}
{Coles} P.~A.,  {Yurchenko} S.~N.,   {Tennyson} J.,  2019, \mn@doi [\mnras]
  {10.1093/mnras/stz2778}, \href
  {https://ui.adsabs.harvard.edu/abs/2019MNRAS.490.4638C} {490, 4638}

\bibitem[\protect\citeauthoryear{{Coxon} \& {Hajigeorgiou}}{{Coxon} \&
  {Hajigeorgiou}}{2015}]{Coxon2015}
{Coxon} J.~A.,  {Hajigeorgiou} P.~G.,  2015, \mn@doi [\jqsrt]
  {10.1016/j.jqsrt.2014.08.028}, \href
  {https://ui.adsabs.harvard.edu/abs/2015JQSRT.151..133C} {151, 133}

\bibitem[\protect\citeauthoryear{{Dalgarno} \& {Williams}}{{Dalgarno} \&
  {Williams}}{1962}]{Dalgarno1962}
{Dalgarno} A.,  {Williams} D.~A.,  1962, \mn@doi [\apj] {10.1086/147428}, \href
  {http://adsabs.harvard.edu/abs/1962ApJ...136..690D} {136, 690}

\bibitem[\protect\citeauthoryear{{Gaia Collaboration} et~al.,}{{Gaia
  Collaboration} et~al.}{2021}]{Gaia3}
{Gaia Collaboration} et~al., 2021, \mn@doi [\aap]
  {10.1051/0004-6361/202039657}, \href
  {https://ui.adsabs.harvard.edu/abs/2021A&A...649A...1G} {649, A1}

\bibitem[\protect\citeauthoryear{{GharibNezhad}, {Shayesteh}  \&
  {Bernath}}{{GharibNezhad} et~al.}{2013}]{GharibNezhad2013}
{GharibNezhad} E.,  {Shayesteh} A.,   {Bernath} P.~F.,  2013, \mn@doi [\mnras]
  {10.1093/mnras/stt510}, \href
  {https://ui.adsabs.harvard.edu/abs/2013MNRAS.432.2043G} {432, 2043}

\bibitem[\protect\citeauthoryear{Gordon et~al.,}{Gordon
  et~al.}{2017}]{Gordon2017}
Gordon I.,  et~al., 2017, \mn@doi [Journal of Quantitative Spectroscopy and
  Radiative Transfer] {https://doi.org/10.1016/j.jqsrt.2017.06.038}, 203, 3

\bibitem[\protect\citeauthoryear{{Gorman}, {Yurchenko}  \& {Tennyson}}{{Gorman}
  et~al.}{2019}]{Gorman2019}
{Gorman} M.~N.,  {Yurchenko} S.~N.,   {Tennyson} J.,  2019, \mn@doi [\mnras]
  {10.1093/mnras/stz2517}, \href
  {https://ui.adsabs.harvard.edu/abs/2019MNRAS.490.1652G} {490, 1652}

\bibitem[\protect\citeauthoryear{{Hargreaves}, {Gordon}, {Kochanov}  \&
  {Rothman}}{{Hargreaves} et~al.}{2019}]{Hargreaves2019}
{Hargreaves} R.,  {Gordon} I.,  {Kochanov} R.,   {Rothman} L.,  2019, in
  EPSC-DPS Joint Meeting 2019. pp EPSC--DPS2019--919

\bibitem[\protect\citeauthoryear{{Hargreaves}, {Gordon}, {Rey}, {Nikitin},
  {Tyuterev}, {Kochanov}  \& {Rothman}}{{Hargreaves}
  et~al.}{2020}]{Hargreaves2020}
{Hargreaves} R.~J.,  {Gordon} I.~E.,  {Rey} M.,  {Nikitin} A.~V.,  {Tyuterev}
  V.~G.,  {Kochanov} R.~V.,   {Rothman} L.~S.,  2020, \mn@doi [\apjs]
  {10.3847/1538-4365/ab7a1a}, \href
  {https://ui.adsabs.harvard.edu/abs/2020ApJS..247...55H} {247, 55}

\bibitem[\protect\citeauthoryear{{Harris}, {Tennyson}, {Kaminsky}, {Pavlenko}
  \& {Jones}}{{Harris} et~al.}{2006}]{Harris2006}
{Harris} G.~J.,  {Tennyson} J.,  {Kaminsky} B.~M.,  {Pavlenko} Y.~V.,   {Jones}
  H.~R.~A.,  2006, \mn@doi [\mnras] {10.1111/j.1365-2966.2005.09960.x}, \href
  {https://ui.adsabs.harvard.edu/abs/2006MNRAS.367..400H} {367, 400}

\bibitem[\protect\citeauthoryear{{Helling} et~al.,}{{Helling}
  et~al.}{2021}]{Helling2021}
{Helling} C.,  et~al., 2021, \mn@doi [\aap] {10.1051/0004-6361/202039911},
  \href {https://ui.adsabs.harvard.edu/abs/2021A&A...649A..44H} {649, A44}

\bibitem[\protect\citeauthoryear{{Heng}}{{Heng}}{2017}]{Heng2017b}
{Heng} K.,  2017, {Exoplanetary Atmospheres: Theoretical Concepts and
  Foundations}

\bibitem[\protect\citeauthoryear{{Hennicker}, {Puls}, {Kee}  \&
  {Sundqvist}}{{Hennicker} et~al.}{2020}]{Hennicker2020}
{Hennicker} L.,  {Puls} J.,  {Kee} N.~D.,   {Sundqvist} J.~O.,  2020, \mn@doi
  [\aap] {10.1051/0004-6361/201936584}, \href
  {https://ui.adsabs.harvard.edu/abs/2020A&A...633A..16H} {633, A16}

\bibitem[\protect\citeauthoryear{Hunter}{Hunter}{2007}]{Hunter2007}
Hunter J.~D.,  2007, \mn@doi [Computing In Science \& Engineering]
  {10.1109/MCSE.2007.55}, 9, 90

\bibitem[\protect\citeauthoryear{{John}}{{John}}{1988}]{John1988}
{John} T.~L.,  1988, \aap, \href
  {https://ui.adsabs.harvard.edu/abs/1988A&A...193..189J} {193, 189}

\bibitem[\protect\citeauthoryear{Jones, Oliphant, Peterson  et~al.}{Jones
  et~al.}{2001}]{Jones2001}
Jones E.,  Oliphant T.,  Peterson P.,   et~al., 2001, {SciPy}: Open source
  scientific tools for {Python}, \url {http://www.scipy.org/}

\bibitem[\protect\citeauthoryear{{Karman} et~al.,}{{Karman}
  et~al.}{2019}]{Karman2019}
{Karman} T.,  et~al., 2019, \mn@doi [\icarus] {10.1016/j.icarus.2019.02.034},
  \href {https://ui.adsabs.harvard.edu/abs/2019Icar..328..160K} {328, 160}

\bibitem[\protect\citeauthoryear{{Kataria}, {Showman}, {Fortney}, {Marley}  \&
  {Freedman}}{{Kataria} et~al.}{2014}]{Kataria2014}
{Kataria} T.,  {Showman} A.~P.,  {Fortney} J.~J.,  {Marley} M.~S.,   {Freedman}
  R.~S.,  2014, \mn@doi [\apj] {10.1088/0004-637X/785/2/92}, \href
  {http://adsabs.harvard.edu/abs/2014ApJ...785...92K} {785, 92}

\bibitem[\protect\citeauthoryear{{Komacek} \& {Showman}}{{Komacek} \&
  {Showman}}{2016}]{Komacek2016}
{Komacek} T.~D.,  {Showman} A.~P.,  2016, \mn@doi [\apj]
  {10.3847/0004-637X/821/1/16}, \href
  {https://ui.adsabs.harvard.edu/abs/2016ApJ...821...16K} {821, 16}

\bibitem[\protect\citeauthoryear{{Komacek}, {Showman}  \& {Tan}}{{Komacek}
  et~al.}{2017}]{Komacek2017}
{Komacek} T.~D.,  {Showman} A.~P.,   {Tan} X.,  2017, \mn@doi [\apj]
  {10.3847/1538-4357/835/2/198}, \href
  {https://ui.adsabs.harvard.edu/abs/2017ApJ...835..198K} {835, 198}

\bibitem[\protect\citeauthoryear{{Kurucz}}{{Kurucz}}{1970}]{Kurucz1970}
{Kurucz} R.~L.,  1970, SAO Special Report, \href
  {https://ui.adsabs.harvard.edu/abs/1970SAOSR.309.....K} {309}

\bibitem[\protect\citeauthoryear{{Kurucz} \& {Bell}}{{Kurucz} \&
  {Bell}}{1995}]{Kurucz1995}
{Kurucz} R.~L.,  {Bell} B.,  1995, {Atomic line list}

\bibitem[\protect\citeauthoryear{{Lee}, {Casewell}, {Chubb}, {Hammond}, {Tan},
  {Tsai}  \& {Pierrehumbert}}{{Lee} et~al.}{2020}]{Lee2020}
{Lee} E. K.~H.,  {Casewell} S.~L.,  {Chubb} K.~L.,  {Hammond} M.,  {Tan} X.,
  {Tsai} S.-M.,   {Pierrehumbert} R.~T.,  2020, \mn@doi [\mnras]
  {10.1093/mnras/staa1882}, \href
  {https://ui.adsabs.harvard.edu/abs/2020MNRAS.496.4674L} {496, 4674}

\bibitem[\protect\citeauthoryear{{Lee} et~al.,}{{Lee} et~al.}{2021a}]{Lee2021b}
{Lee} E. K.~H.,  et~al., 2021a, arXiv e-prints, \href
  {https://ui.adsabs.harvard.edu/abs/2021arXiv211015640L} {p. arXiv:2110.15640}

\bibitem[\protect\citeauthoryear{{Lee}, {Parmentier}, {Hammond}, {Grimm},
  {Kitzmann}, {Tan}, {Tsai}  \& {Pierrehumbert}}{{Lee} et~al.}{2021b}]{Lee2021}
{Lee} E. K.~H.,  {Parmentier} V.,  {Hammond} M.,  {Grimm} S.~L.,  {Kitzmann}
  D.,  {Tan} X.,  {Tsai} S.-M.,   {Pierrehumbert} R.~T.,  2021b, \mn@doi
  [\mnras] {10.1093/mnras/stab1851}, \href
  {https://ui.adsabs.harvard.edu/abs/2021MNRAS.506.2695L} {506, 2695}

\bibitem[\protect\citeauthoryear{{Lew} et~al.,}{{Lew} et~al.}{2022}]{Lew2022}
{Lew} B. W.~P.,  et~al., 2022, \mn@doi [\aj] {10.3847/1538-3881/ac3001}, \href
  {https://ui.adsabs.harvard.edu/abs/2022AJ....163....8L} {163, 8}

\bibitem[\protect\citeauthoryear{{Li}, {Gordon}, {Le Roy}, {Hajigeorgiou},
  {Coxon}, {Bernath}  \& {Rothman}}{{Li} et~al.}{2013}]{Li2013}
{Li} G.,  {Gordon} I.~E.,  {Le Roy} R.~J.,  {Hajigeorgiou} P.~G.,  {Coxon}
  J.~A.,  {Bernath} P.~F.,   {Rothman} L.~S.,  2013, \mn@doi [\jqsrt]
  {10.1016/j.jqsrt.2013.02.005}, \href
  {https://ui.adsabs.harvard.edu/abs/2013JQSRT.121...78L} {121, 78}

\bibitem[\protect\citeauthoryear{{Li}, {Gordon}, {Rothman}, {Tan}, {Hu},
  {Kassi}, {Campargue}  \& {Medvedev}}{{Li} et~al.}{2015}]{Li2015}
{Li} G.,  {Gordon} I.~E.,  {Rothman} L.~S.,  {Tan} Y.,  {Hu} S.-M.,  {Kassi}
  S.,  {Campargue} A.,   {Medvedev} E.~S.,  2015, \mn@doi [The Astrophysical
  Journal Supplement Series] {10.1088/0067-0049/216/1/15}, \href
  {https://ui.adsabs.harvard.edu/#abs/2015ApJS..216...15L} {216, 15}

\bibitem[\protect\citeauthoryear{{Littlefair} et~al.,}{{Littlefair}
  et~al.}{2014}]{Littlefair2014}
{Littlefair} S.~P.,  et~al., 2014, \mn@doi [\mnras] {10.1093/mnras/stu1895},
  \href {https://ui.adsabs.harvard.edu/#abs/2014MNRAS.445.2106L} {445, 2106}

\bibitem[\protect\citeauthoryear{{Liu} \& {Showman}}{{Liu} \&
  {Showman}}{2013}]{Liu2013}
{Liu} B.,  {Showman} A.~P.,  2013, \mn@doi [\apj] {10.1088/0004-637X/770/1/42},
  \href {https://ui.adsabs.harvard.edu/abs/2013ApJ...770...42L} {770, 42}

\bibitem[\protect\citeauthoryear{{Longstaff}, {Casewell}, {Wynn}, {Maxted}  \&
  {Helling}}{{Longstaff} et~al.}{2017}]{Longstaff2017}
{Longstaff} E.~S.,  {Casewell} S.~L.,  {Wynn} G.~A.,  {Maxted} P.~F.~L.,
  {Helling} C.,  2017, \mn@doi [\mnras] {10.1093/mnras/stx1786}, \href
  {https://ui.adsabs.harvard.edu/abs/2017MNRAS.471.1728L} {471, 1728}

\bibitem[\protect\citeauthoryear{{Lothringer} \& {Casewell}}{{Lothringer} \&
  {Casewell}}{2020}]{Lothringer2020}
{Lothringer} J.~D.,  {Casewell} S.~L.,  2020, \mn@doi [\apj]
  {10.3847/1538-4357/abc5bc}, \href
  {https://ui.adsabs.harvard.edu/abs/2020ApJ...905..163L} {905, 163}

\bibitem[\protect\citeauthoryear{{Lothringer}, {Fu}, {Sing}  \&
  {Barman}}{{Lothringer} et~al.}{2020}]{Lothringer2020b}
{Lothringer} J.~D.,  {Fu} G.,  {Sing} D.~K.,   {Barman} T.~S.,  2020, \mn@doi
  [\apjl] {10.3847/2041-8213/aba265}, \href
  {https://ui.adsabs.harvard.edu/abs/2020ApJ...898L..14L} {898, L14}

\bibitem[\protect\citeauthoryear{{Marley}}{{Marley}}{2000}]{Marley2000}
{Marley} M.,  2000, in {Griffith} C.~A.,  {Marley} M.~S.,  eds,  Astronomical
  Society of the Pacific Conference Series Vol. 212, From Giant Planets to Cool
  Stars. p.~152

\bibitem[\protect\citeauthoryear{{Marley}, {Seager}, {Saumon}, {Lodders},
  {Ackerman}, {Freedman}  \& {Fan}}{{Marley} et~al.}{2002}]{Marley2002}
{Marley} M.~S.,  {Seager} S.,  {Saumon} D.,  {Lodders} K.,  {Ackerman} A.~S.,
  {Freedman} R.~S.,   {Fan} X.,  2002, \mn@doi [\apj] {10.1086/338800}, \href
  {https://ui.adsabs.harvard.edu/abs/2002ApJ...568..335M} {568, 335}

\bibitem[\protect\citeauthoryear{{Maxted}, {Napiwotzki}, {Dobbie}  \&
  {Burleigh}}{{Maxted} et~al.}{2006}]{Maxted2006}
{Maxted} P.~F.~L.,  {Napiwotzki} R.,  {Dobbie} P.~D.,   {Burleigh} M.~R.,
  2006, \mn@doi [\nat] {10.1038/nature04987}, \href
  {https://ui.adsabs.harvard.edu/abs/2006Natur.442..543M} {442, 543}

\bibitem[\protect\citeauthoryear{{McKemmish}, {Yurchenko}  \&
  {Tennyson}}{{McKemmish} et~al.}{2016}]{McKemmish2016}
{McKemmish} L.~K.,  {Yurchenko} S.~N.,   {Tennyson} J.,  2016, \mn@doi [\mnras]
  {10.1093/mnras/stw1969}, \href
  {https://ui.adsabs.harvard.edu/abs/2016MNRAS.463..771M} {463, 771}

\bibitem[\protect\citeauthoryear{{McKemmish}, {Masseron}, {Hoeijmakers},
  {P{\'e}rez-Mesa}, {Grimm}, {Yurchenko}  \& {Tennyson}}{{McKemmish}
  et~al.}{2019}]{McKemmish2019}
{McKemmish} L.~K.,  {Masseron} T.,  {Hoeijmakers} H.~J.,  {P{\'e}rez-Mesa} V.,
  {Grimm} S.~L.,  {Yurchenko} S.~N.,   {Tennyson} J.,  2019, \mn@doi [\mnras]
  {10.1093/mnras/stz1818}, \href
  {https://ui.adsabs.harvard.edu/abs/2019MNRAS.488.2836M} {488, 2836}

\bibitem[\protect\citeauthoryear{{Mendon{\c{c}}a}, {Read}, {Wilson}  \&
  {Lee}}{{Mendon{\c{c}}a} et~al.}{2015}]{Mendonca2015}
{Mendon{\c{c}}a} J.~M.,  {Read} P.~L.,  {Wilson} C.~F.,   {Lee} C.,  2015,
  \mn@doi [\planss] {10.1016/j.pss.2014.11.008}, \href
  {https://ui.adsabs.harvard.edu/abs/2015P&SS..105...80M} {105, 80}

\bibitem[\protect\citeauthoryear{{Morley}, {Fortney}, {Marley}, {Visscher},
  {Saumon}  \& {Leggett}}{{Morley} et~al.}{2012}]{Morley2012}
{Morley} C.~V.,  {Fortney} J.~J.,  {Marley} M.~S.,  {Visscher} C.,  {Saumon}
  D.,   {Leggett} S.~K.,  2012, \mn@doi [\apj] {10.1088/0004-637X/756/2/172},
  \href {http://adsabs.harvard.edu/abs/2012ApJ...756..172M} {756, 172}

\bibitem[\protect\citeauthoryear{{Morley}, {Fortney}, {Kempton}, {Marley},
  {Visscher}  \& {Zahnle}}{{Morley} et~al.}{2013}]{Morley2013}
{Morley} C.~V.,  {Fortney} J.~J.,  {Kempton} E.~M.-R.,  {Marley} M.~S.,
  {Visscher} C.,   {Zahnle} K.,  2013, \mn@doi [\apj]
  {10.1088/0004-637X/775/1/33}, \href
  {http://adsabs.harvard.edu/abs/2013ApJ...775...33M} {775, 33}

\bibitem[\protect\citeauthoryear{{Olson} \& {Kunasz}}{{Olson} \&
  {Kunasz}}{1987}]{Olson1987}
{Olson} G.~L.,  {Kunasz} P.~B.,  1987, \mn@doi [\jqsrt]
  {10.1016/0022-4073(87)90027-6}, \href
  {https://ui.adsabs.harvard.edu/abs/1987JQSRT..38..325O} {38, 325}

\bibitem[\protect\citeauthoryear{{Parmentier}, {Guillot}, {Fortney}  \&
  {Marley}}{{Parmentier} et~al.}{2015}]{Parmentier2015}
{Parmentier} V.,  {Guillot} T.,  {Fortney} J.~J.,   {Marley} M.~S.,  2015,
  \mn@doi [\aap] {10.1051/0004-6361/201323127}, \href
  {https://ui.adsabs.harvard.edu/abs/2015A&A...574A..35P} {574, A35}

\bibitem[\protect\citeauthoryear{{Parmentier}, {Showman}  \&
  {Fortney}}{{Parmentier} et~al.}{2021}]{Parmentier2021}
{Parmentier} V.,  {Showman} A.~P.,   {Fortney} J.~J.,  2021, \mn@doi [\mnras]
  {10.1093/mnras/staa3418}, \href
  {https://ui.adsabs.harvard.edu/abs/2021MNRAS.501...78P} {501, 78}

\bibitem[\protect\citeauthoryear{{Polyansky}, {Kyuberis}, {Zobov}, {Tennyson},
  {Yurchenko}  \& {Lodi}}{{Polyansky} et~al.}{2018}]{Polyansky2018}
{Polyansky} O.~L.,  {Kyuberis} A.~A.,  {Zobov} N.~F.,  {Tennyson} J.,
  {Yurchenko} S.~N.,   {Lodi} L.,  2018, \mn@doi [\mnras]
  {10.1093/mnras/sty1877}, \href
  {http://adsabs.harvard.edu/abs/2018MNRAS.480.2597P} {480, 2597}

\bibitem[\protect\citeauthoryear{{Roueff}, {Abgrall}, {Czachorowski},
  {Pachucki}, {Puchalski}  \& {Komasa}}{{Roueff} et~al.}{2019}]{Roueff2019}
{Roueff} E.,  {Abgrall} H.,  {Czachorowski} P.,  {Pachucki} K.,  {Puchalski}
  M.,   {Komasa} J.,  2019, \mn@doi [\aap] {10.1051/0004-6361/201936249}, \href
  {https://ui.adsabs.harvard.edu/abs/2019A&A...630A..58R} {630, A58}

\bibitem[\protect\citeauthoryear{{Sainsbury-Martinez}, {Casewell},
  {Lothringer}, {Phillips}  \& {Tremblin}}{{Sainsbury-Martinez}
  et~al.}{2021}]{Sainsbury-Martinez2021}
{Sainsbury-Martinez} F.,  {Casewell} S.~L.,  {Lothringer} J.~D.,  {Phillips}
  M.~W.,   {Tremblin} P.,  2021, \mn@doi [\aap] {10.1051/0004-6361/202141637},
  \href {https://ui.adsabs.harvard.edu/abs/2021A&A...656A.128S} {656, A128}

\bibitem[\protect\citeauthoryear{{Sharp} \& {Burrows}}{{Sharp} \&
  {Burrows}}{2007}]{Sharp2007}
{Sharp} C.~M.,  {Burrows} A.,  2007, \mn@doi [\apjs] {10.1086/508708}, \href
  {http://adsabs.harvard.edu/abs/2007ApJS..168..140S} {168, 140}

\bibitem[\protect\citeauthoryear{{Showman}}{{Showman}}{2016}]{Showman2016}
{Showman} A.~P.,  2016, \mn@doi [\nat] {10.1038/533330a}, \href
  {https://ui.adsabs.harvard.edu/abs/2016Natur.533..330S} {533, 330}

\bibitem[\protect\citeauthoryear{{Showman}, {Fortney}, {Lian}, {Marley},
  {Freedman}, {Knutson}  \& {Charbonneau}}{{Showman}
  et~al.}{2009}]{Showman2009}
{Showman} A.~P.,  {Fortney} J.~J.,  {Lian} Y.,  {Marley} M.~S.,  {Freedman}
  R.~S.,  {Knutson} H.~A.,   {Charbonneau} D.,  2009, \mn@doi [\apj]
  {10.1088/0004-637X/699/1/564}, \href
  {http://adsabs.harvard.edu/abs/2009ApJ...699..564S} {699, 564}

\bibitem[\protect\citeauthoryear{{Skrutskie} et~al.,}{{Skrutskie}
  et~al.}{2006}]{Skrutskie2006}
{Skrutskie} M.~F.,  et~al., 2006, \mn@doi [\aj] {10.1086/498708}, \href
  {https://ui.adsabs.harvard.edu/abs/2006AJ....131.1163S} {131, 1163}

\bibitem[\protect\citeauthoryear{{Sousa-Silva}, {Al-Refaie}, {Tennyson}  \&
  {Yurchenko}}{{Sousa-Silva} et~al.}{2015}]{Sousa_Silva2015}
{Sousa-Silva} C.,  {Al-Refaie} A.~F.,  {Tennyson} J.,   {Yurchenko} S.~N.,
  2015, \mn@doi [\mnras] {10.1093/mnras/stu2246}, \href
  {https://ui.adsabs.harvard.edu/abs/2015MNRAS.446.2337S} {446, 2337}

\bibitem[\protect\citeauthoryear{{Steele}, {Burleigh}, {Dobbie}, {Jameson},
  {Barstow}  \& {Satterthwaite}}{{Steele} et~al.}{2011}]{Steele2011}
{Steele} P.~R.,  {Burleigh} M.~R.,  {Dobbie} P.~D.,  {Jameson} R.~F.,
  {Barstow} M.~A.,   {Satterthwaite} R.~P.,  2011, \mn@doi [\mnras]
  {10.1111/j.1365-2966.2011.19225.x}, \href
  {https://ui.adsabs.harvard.edu/abs/2011MNRAS.416.2768S} {416, 2768}

\bibitem[\protect\citeauthoryear{{Steinrueck}, {Showman}, {Lavvas}, {Koskinen},
  {Tan}  \& {Zhang}}{{Steinrueck} et~al.}{2021}]{Steinrueck2021}
{Steinrueck} M.~E.,  {Showman} A.~P.,  {Lavvas} P.,  {Koskinen} T.,  {Tan} X.,
   {Zhang} X.,  2021, \mn@doi [\mnras] {10.1093/mnras/stab1053}, \href
  {https://ui.adsabs.harvard.edu/abs/2021MNRAS.504.2783S} {504, 2783}

\bibitem[\protect\citeauthoryear{{Tan} \& {Komacek}}{{Tan} \&
  {Komacek}}{2019}]{Tan2019}
{Tan} X.,  {Komacek} T.~D.,  2019, \mn@doi [\apj] {10.3847/1538-4357/ab4a76},
  \href {https://ui.adsabs.harvard.edu/abs/2019ApJ...886...26T} {886, 26}

\bibitem[\protect\citeauthoryear{{Tan} \& {Showman}}{{Tan} \&
  {Showman}}{2020}]{Tan2020}
{Tan} X.,  {Showman} A.~P.,  2020, \mn@doi [\apj] {10.3847/1538-4357/abb3d4},
  \href {https://ui.adsabs.harvard.edu/abs/2020ApJ...902...27T} {902, 27}

\bibitem[\protect\citeauthoryear{{Thalman}, {Zarzana}, {Tolbert}  \&
  {Volkamer}}{{Thalman} et~al.}{2014}]{Thalman2014}
{Thalman} R.,  {Zarzana} K.~J.,  {Tolbert} M.~A.,   {Volkamer} R.,  2014,
  \mn@doi [\jqsrt] {10.1016/j.jqsrt.2014.05.030}, \href
  {https://ui.adsabs.harvard.edu/abs/2014JQSRT.147..171T} {147, 171}

\bibitem[\protect\citeauthoryear{{The Astropy Collaboration} et~al.,}{{The
  Astropy Collaboration} et~al.}{2018}]{Astropy2018}
{The Astropy Collaboration} et~al., 2018, preprint, \href
  {http://adsabs.harvard.edu/abs/2018arXiv180102634T} {} (\mn@eprint {arXiv}
  {1801.02634})

\bibitem[\protect\citeauthoryear{{Vanderburg} et~al.,}{{Vanderburg}
  et~al.}{2020}]{Vanderburg2020}
{Vanderburg} A.,  et~al., 2020, \mn@doi [\nat] {10.1038/s41586-020-2713-y},
  \href {https://ui.adsabs.harvard.edu/abs/2020Natur.585..363V} {585, 363}

\bibitem[\protect\citeauthoryear{{Wakeford}, {Visscher}, {Lewis}, {Kataria},
  {Marley}, {Fortney}  \& {Mandell}}{{Wakeford} et~al.}{2017}]{Wakeford2017a}
{Wakeford} H.~R.,  {Visscher} C.,  {Lewis} N.~K.,  {Kataria} T.,  {Marley}
  M.~S.,  {Fortney} J.~J.,   {Mandell} A.~M.,  2017, \mn@doi [\mnras]
  {10.1093/mnras/stw2639}, \href
  {http://adsabs.harvard.edu/abs/2017MNRAS.464.4247W} {464, 4247}

\bibitem[\protect\citeauthoryear{{Western}, {Carter-Blatchford}, {Crozet},
  {Ross}, {Morville}  \& {Tokaryk}}{{Western} et~al.}{2018}]{Western2018}
{Western} C.~M.,  {Carter-Blatchford} L.,  {Crozet} P.,  {Ross} A.~J.,
  {Morville} J.,   {Tokaryk} D.~W.,  2018, \mn@doi [\jqsrt]
  {10.1016/j.jqsrt.2018.07.017}, \href
  {https://ui.adsabs.harvard.edu/abs/2018JQSRT.219..127W} {219, 127}

\bibitem[\protect\citeauthoryear{{Woitke}, {Helling}, {Hunter}, {Millard},
  {Turner}, {Worters}, {Blecic}  \& {Stock}}{{Woitke}
  et~al.}{2018}]{Woitke2018}
{Woitke} P.,  {Helling} C.,  {Hunter} G.~H.,  {Millard} J.~D.,  {Turner} G.~E.,
   {Worters} M.,  {Blecic} J.,   {Stock} J.~W.,  2018, \mn@doi [\aap]
  {10.1051/0004-6361/201732193}, \href
  {https://ui.adsabs.harvard.edu/#abs/2018A&A...614A...1W} {614, A1}

\bibitem[\protect\citeauthoryear{{Yurchenko}, {Mellor}, {Freedman}  \&
  {Tennyson}}{{Yurchenko} et~al.}{2020}]{Yurchenko2020}
{Yurchenko} S.~N.,  {Mellor} T.~M.,  {Freedman} R.~S.,   {Tennyson} J.,  2020,
  \mn@doi [\mnras] {10.1093/mnras/staa1874}, \href
  {https://ui.adsabs.harvard.edu/abs/2020MNRAS.496.5282Y} {496, 5282}

\bibitem[\protect\citeauthoryear{{Yurchenko} et~al.,}{{Yurchenko}
  et~al.}{2021}]{Yurchenko2021}
{Yurchenko} S.~N.,  et~al., 2021, \mn@doi [\mnras] {10.1093/mnras/stab3267},
  \href {https://ui.adsabs.harvard.edu/abs/2021MNRAS.tmp.2980Y} {}

\bibitem[\protect\citeauthoryear{{Zhou} et~al.,}{{Zhou}
  et~al.}{2021}]{Zhou2021}
{Zhou} Y.,  et~al., 2021, arXiv e-prints, \href
  {https://ui.adsabs.harvard.edu/abs/2021arXiv211010162Z} {p. arXiv:2110.10162}

\bibitem[\protect\citeauthoryear{{Zhou} et~al.,}{{Zhou}
  et~al.}{2022}]{Zhou2022}
{Zhou} Y.,  et~al., 2022, \mn@doi [\aj] {10.3847/1538-3881/ac3095}, \href
  {https://ui.adsabs.harvard.edu/abs/2022AJ....163...17Z} {163, 17}

\makeatother
\end{thebibliography}



\clearpage

\appendix


\bsp	
\label{lastpage}
\end{document}